\documentclass[11pt]{article}
\baselineskip
\usepackage{amssymb, graphicx, latexsym, cite, epsfig, amsmath, bbm, color}
\usepackage[vcentermath]{youngtab}
\usepackage{hyperref}

\newcommand{\Z}{\mathbb{Z}}

\newcommand{\SU}{\mathrm{SU}}
\newcommand{\U}{\mathrm{U}}

\newcommand{\dd}{{\rm{d}}}
\newcommand{\Tr}{{\rm Tr\,}}
\newcommand{\real}{{\rm Re\,}}

\newcommand{\Sel}{S^{\tiny\mbox{E}}_{\tiny\mbox{L}}}
\newcommand{\Lren}{L^{\mbox{\tiny{ren}}}}
\newcommand{\Fren}{F^{\mbox{\tiny{ren}}}}

\newcommand{\nconf}{n_{\mbox{\tiny{conf}}}}
\newcommand{\redchisq}{\chi^2_{\tiny\mbox{red}}}
\newcommand{\eq}{\begin{equation}}
\newcommand{\en}{\end{equation}}
\newcommand{\eqar}{\begin{eqnarray}}
\newcommand{\enar}{\end{eqnarray}}

\begin{document}

\begin{titlepage}
\vskip0.5cm 
\begin{flushright} 
HIP-2012-05/TH\\ 
NSF-KITP-12-013
\end{flushright} 
\vskip0.5cm 
\begin{center}
{\Large\bf
Casimir scaling and renormalization of Polyakov loops in large-$N$ gauge theories
}
\end{center}
\vskip1.3cm
\centerline{Anne~Mykk\"anen$^{a}$, Marco~Panero$^{a,b}$ and Kari~Rummukainen$^{a}$}
\vskip1.5cm
\centerline{\sl $^a$ Department of Physics and Helsinki Institute of Physics}
\centerline{\sl P.O. Box 64, FI-00014 University of Helsinki, Finland}
\vskip0.5cm
\centerline{\sl $^b$ Kavli Institute for Theoretical Physics}
\centerline{\sl University of California, Santa Barbara, CA 93106, USA}
\vskip0.5cm
\begin{center}
{\sl  E-mail:} \hskip 5mm \texttt{anne-mari.mykkanen@helsinki.fi, marco.panero@helsinki.fi, kari.rummukainen@helsinki.fi}
\end{center}
\vskip1.0cm
\begin{abstract}
We study Casimir scaling and renormalization properties of Polyakov loops in different irreducible representations in $\SU(N)$ gauge theories; in particular, we investigate the approach to the large-$N$ limit, by performing lattice simulations of Yang-Mills theories with an increasing number of colors, from $2$ to $6$. We consider the twelve lowest irreducible representations for each gauge group, and find strong numerical evidence for nearly perfect Casimir scaling of the bare Polyakov loops in the deconfined phase. Then we discuss the temperature dependence of renormalized loops, which is found to be qualitatively and quantitatively very similar for the various gauge groups. In particular, close to the deconfinement transition, the renormalized Polyakov loop increases with the temperature, and its logarithm reveals a characteristic dependence on the inverse of the square of the temperature. At higher temperatures, the renormalized Polyakov loop overshoots one, reaches a maximum, and then starts decreasing, in agreement with weak-coupling predictions. The implications of these findings are discussed.
\end{abstract}
\vspace*{0.2cm}

\end{titlepage}

\section{Introduction and motivation} 
\label{sec:intro} 

The change of state to a deconfined phase at high temperatures or densities is a very important phenomenon in quantum chromodynamics (QCD) and in other non-Abelian gauge theories. While at zero and low temperatures the physical states are color-singlet hadronic states, in the high-temperature limit the physical running coupling becomes small, due to asymptotic freedom, and one expects that the physics should be described in terms of a gas of weakly interacting quarks and gluons~\cite{first_deconfinement_prediction}: the quark-gluon plasma (QGP)~\cite{Shuryak:1978ij}. These two qualitatively different phases should be separated by a phase transition or a crossover, which has been searched for in an extensive experimental heavy-ion collision programme since the 1980's. The results obtained at SPS, RHIC and LHC during the last decade show, indeed, convincing evidence for the creation of a new state of matter at temperatures about 160~MeV, which behaves as an almost ideal fluid~\cite{finite_T_experiments}. The experimental research on the QCD phase diagram will be continued and extended at FAIR and NICA.

On the theoretical side, however, the quantitative understanding of the QCD plasma is still an open problem. One of the reasons for this is that the deconfined plasma retains some non-perturbative features even in the limit of high temperatures $T$. In particular, the presence of severe infrared divergences in weak-coupling expansions for thermal gauge theories leads to non-analytical properties of the perturbative series for various physical observables, and to a breakdown of the correspondence between loop expansions and expansions in powers of the coupling~\cite{infrared_divergences}. As a consequence, the long-wavelength modes of the QGP are strongly coupled at all temperatures, and thus cannot be treated perturbatively---see ref.~\cite{Blaizot:2011va} for a review. Finally, at the typical temperatures probed in experiments, the physical coupling of QCD turns out to be relatively small, but not extremely so, and perturbative predictions fail close to the deconfinement temperature~\cite{perturbative_EoS}.

For these reasons, the theoretical study of the QGP at temperatures close to the deconfining transition is usually addressed with non-perturbative methods, including, in particular, numerical simulations on the lattice~\cite{finite_T_lattice_reviews}. During the last decade, lattice computations of the equation of state in QCD with light dynamical quarks have reached high levels of precision, and showed that the deconfinement at finite temperature and vanishing quark chemical potential (for physical values of the quark masses) is a crossover, rather than a genuine phase transition. In fact, in QCD with quarks of finite mass there is no exact symmetry-breaking pattern to characterize the deconfinement.

By contrast, pure $\SU(N)$ Yang-Mills theories (which capture most of the qualitative features of the physics of deconfinement) provide much a cleaner theoretical setup: in the Euclidean formulation, it is easy to see that the Lagrangian of $\SU(N)$ Yang-Mills theories at finite temperature is invariant under a global symmetry associated with the center of the gauge group $\Z_N$~\cite{tHooft:1977hy}. The order parameter for this symmetry is the trace of the temporal Wilson line, or Polyakov loop~\cite{old_Polyakov_loop}:
\begin{equation}
\label{bare_loop_definition}
L = \langle \Tr L(\vec{x}) \rangle = \left\langle \Tr \mathcal{P} \exp \left[ ig_0 \int_0^{1/T} \!\!\!\!\! \dd \tau A_0 (\tau, \vec{x}) \right] \right\rangle.
\end{equation}
In the thermodynamic limit, the ground-state expectation value of $L$ is exactly vanishing in the low-temperature phase, while it becomes non-zero above the critical deconfinement temperature $T_c$, signaling the spontaneous breakdown of center symmetry. Although $L$, \emph{per se}, is not a physical observable, it can be interpreted as the trace of the propagator of an external, infinitely massive probe color charge located at $\vec{x}$: a vanishing $L$ in the $\Z_N$-symmetric ground state at low temperatures $T<T_c$ means that the expectation value of a static color charge is zero, and hence the system is confined. On the contrary, $L$ is non-zero in the high-temperature phase at $T> T_c$, corresponding to a finite free energy for the probe color charge in the deconfined phase. Thus, $L$ has the meaning of an order parameter for the finite-temperature deconfinement transition in Yang-Mills theory. Another possible order parameter for the transition is given by the two-point Polyakov loop correlation function: across the phase transition, it changes from confining to exponentially screened. The Polyakov loop correlation function extracted from lattice simulations at finite temperature is often used as an input for effective potential models for quarkonia~\cite{quarkonium_potential_models}; however, certain subtleties related to the connection between the real- and the imaginary-time formalism, and to the spectral decomposition into singlet and octet contributions to the corresponding free energies have recently been pointed out in the literature~\cite{Jahn_Philipsen}.

Note that the free energy associated with the \emph{bare} Polyakov loop defined by eq.~(\ref{bare_loop_definition}) is a divergent quantity, and hence needs to be renormalized~\cite{Dotsenko:1979wb}.

In general, in $\SU(N)$ Yang-Mills theory the Polyakov loop is an order parameter for a probe charge in a generic irreducible representation of the gauge group with non-zero $N$-ality (i.e., a representation transforming non-trivially under the center of the group). The free energy associated with charges in different irreducible representations is expected to be proportional to the eigenvalue of the corresponding quadratic Casimir operator $\langle C_2 \rangle$~\cite{Ambjorn:1984mb}. This property is called ``Casimir scaling'': it is not specific to Polyakov loops, and indeed it has been studied for various other observables~\cite{Casimir_scaling_references} (see also ref.~\cite{Shevchenko_Simonov} for a discussion). For the Polyakov loop, perturbative calculations predict Casimir scaling to hold at the lowest orders~\cite{perturbative_Polyakov_loop} (deviations from Casimir scaling are predicted to occur only at $O(g^6)$). 

In this work, we study the behavior of bare and renormalized Polyakov loops in non-Abelian gauge theories with a different number of colors, from $2$ to $6$, discussing various renormalization methods, and comparing our results to those of recent, similar studies for $\SU(3)$~\cite{Kaczmarek:2002mc, Dumitru:2003hp, Gupta:2007ax} and $\SU(2)$~\cite{Digal:2003jc, Hubner:2008ef, Gavai:2010qd} Yang-Mills theories. In particular, we investigate the features that emerge when $N$ is large. The motivations for looking at the limit of a large number of colors are manifold. First of all, the large-$N$ limit of QCD at fixed 't~Hooft coupling $\lambda=g^2 N$ and fixed number of flavors $N_f$ is known to lead to dramatic mathematical simplifications~\cite{largeN}. For the phase diagram of QCD-like theories, the large-$N$ limit has also interesting implications for new phases at high density~\cite{quarkyonic_matter}. Furthermore, it plays a technically crucial r\^ole in holographic computations, inspired by the conjectured equivalence of maximally supersymmetric Yang-Mills theory with $\mathcal{N}=4$ supercharges in four dimensions and supersymmetric type IIB string theory in a $10$-dimensional $AdS_5 \times S^5$ spacetime~\cite{Maldacena_conjecture}. This conjecture relates the large-$N$ limit of the strongly coupled gauge theory to the classical gravity limit of string theory in a five-dimensional anti-de~Sitter spacetime, which can be studied analytically. While at zero temperature the $\mathcal{N}=4$ theory is qualitatively very different from QCD, there are arguments suggesting that at finite temperature the two theories should share at least some qualitative (or semi-quantitative) physical features~\cite{Witten:1998zw}. Calculations based on the gauge/string duality have also been extended to various other models, which mimic the features of QCD either by breaking explicitly some of the symmetries of the $\mathcal{N}=4$ theory using some additional ingredients (``top-down'' approach), or by constructing some \emph{ad hoc} five-dimensional gravity model, which should reproduce the properties of QCD (``bottom-up'' approach). These models are often used to study analytically certain features of the strongly coupled quark-gluon plasma~\cite{gauge_string_applications_to_QCD_plasma}.

One important technical aspect in all holographic computations is that they are based on the approximation of an infinite number of colors in the gauge theory: this limit allows one to neglect loop effects in the dual string theory, i.e. to reduce it to its classical limit. Recent lattice studies have showed that the large-$N$ limit is indeed a good approximation for the physical $\SU(3)$ case, both as it concerns spectral and thermal observables~\cite{largeN3+1D}; remarkably, this also holds for theories in $2+1$ spacetime dimensions~\cite{largeN2+1D}. However, the validity of the infinite-$N$ approximation is, in general, a non-trivial issue, which can depend on the observable considered, and should be studied on a case-by-case basis.

In the context of gauge/string duality, the behavior of the renormalized Polyakov loop as a function of the temperature has been recently discussed in refs.~\cite{Noronha:2009ud, Andreev:2009zk, Megias:2010ku}. In particular, in ref.~\cite{Noronha:2009ud} it was argued that, in strongly coupled theories with a holographic dual, the renormalized Polyakov loop should be monotonically increasing with $T$. This is in contrast with perturbative computations~\cite{perturbative_Polyakov_loop}, which predict that the leading-order correction to the free limit is positive, 
and hence that the renormalized loop $\Lren$ should tend to unity \emph{from above} in the high-temperature limit. However, it should be noted that these two theoretical predictions are expected to hold in the strong- and in the weak-coupling regime, respectively. A holographic prediction for the renormalized Polyakov loop was worked out analytically in ref.~\cite{Andreev:2009zk}, using a simple holographic model with one deformation parameter~\cite{Andreev:2007zv}. This work found that, at the leading order in a high-temperature expansion, the logarithm of the Polyakov loop in the strong coupling regime should be given by the sum of a constant plus a term proportional to $(T_c/T)^2$, an effect which has also been observed and discussed in refs.~\cite{Megias:2005ve, Megias_Ruiz_Arriola_Salcedo, Xu:2011ud}.

The properties of renormalized Polyakov loops in theories based on different gauge groups are also of interest for effective models of the quark-gluon plasma in the region near $T_c$, see refs.~\cite{Meisinger:2001cq, Polyakov_loop_model, Dumitru:2003hp} and references therein. In particular, the behavior in the large-$N$ limit may reveal analogies with the third-order transition that one finds in $1+1$ dimensions~\cite{Gross_Witten_Wadia}. Moreover, at large $N$ one expects that different irreducible representations become equivalent, up to $O(1/N)$ corrections: for example, the two-index symmetric and antisymmetric representations are expected to be equivalent for $N \to \infty$. Furthermore, using the group theoretical tools of composite representations~\cite{Gross_Taylor} (see the appendix~\ref{sec:appendix} for details), it is possible to show that in the large-$N$ limit the eigenvalue of the quadratic Casimir remains $O(N)$.

Finally, the finite-temperature properties of strongly coupled gauge theories based on different gauge groups and with \emph{dynamical} fermions in various representations are also interesting for extended technicolor models~\cite{ETC}.

With this motivation, in this work we address a first-principle lattice study of Polyakov loops at finite temperature in $\SU(N)$ gauge theories with a different number of colors $N$, and for several irreducible representations. In particular, we consider the twelve lowest non-trivial irreducible representations of each gauge group, and investigate the Casimir scaling at temperatures close to the deconfinement transition. Then we define non-perturbatively renormalized Polyakov loops, discussing various renormalization methods that have recently been proposed in the literature. While all our computations are performed in the setup of the pure Yang-Mills theory, it is worth remarking that, in the 't~Hooft limit, the dynamics of gluons dominates, with the contributions from virtual quark loops suppressed by powers of $1/N$: the large-$N$ limit of QCD is a unitary quenched theory, and by virtue of this, in this limit it is legitimate to consider only the glue sector of the theory on the lattice. This allows one to avoid the complications arising from lattice fermions, and to achieve a smoother approach to the planar limit (the leading-order finite-$N$ corrections in the glue sector are proportional to $1/N^2$).

In section~\ref{sec:lattice} we define the setup of our lattice computations and the method to extract the renormalized Polyakov loop free energies. Our results are presented in section~\ref{sec:results}, while in  section~\ref{sec:conclusions} we discuss their implications, and summarize our findings. Some useful group-theoretical formul{\ae} are listed in the appendix~\ref{sec:appendix}.

Preliminary results of this study were presented in ref.~\cite{Mykkanen:2011kz}.

\section{Lattice simulation setup}
\label{sec:lattice}

Our numerical simulations are based on the regularization of $\SU(N)$ Yang-Mills theories with $N=2$, $3$, $4$, $5$ and $6$ colors on a four-dimensional Euclidean hypercubic, isotropic lattice $\Lambda$ of spacing $a$, with periodic boundary conditions in all directions. We use natural units ($\hbar=c=k_{\tiny{\mbox{B}}}=1$), so that the temperature equals the inverse of the size of the system in the compactified Euclidean time direction: $T=1/(aN_t)$, and denote the spatial volume of the lattice as $V=(aN_s)^3$. For most of our simulations at finite temperature, we used lattices characterized by an aspect ratio $N_s/N_t \ge 4$, which provides a good approximation of the thermodynamic limit~\cite{finite_volume_finite_T}. The fundamental degrees of freedom in the lattice regularization of the theory are a discrete (and finite, if one considers a finite hypervolume) set of $U_\mu(x)$ matrices in the $N \times N$ representation of the group, which are defined on (and represent parallel transporters along) the oriented bonds between nearest-neighbor sites on the lattice. The functional integral defining the continuum partition function of the system is traded for a well-defined, finite, multi-dimensional ordinary integral:
\eq
\label{lattice_partition_function}
Z = \int \prod_{x \in \Lambda} \prod_{\alpha=1}^4 {\dd}U_\alpha(x) e^{-\Sel},
\en
where ${\dd}U_\alpha(x)$ is the Haar measure for each $U_\alpha(x) \in \SU(N)$ link matrix, and $\Sel$ denotes a gauge-invariant lattice action. The simplest choice for $\Sel$ is given by the Wilson gauge action~\cite{Wilson:1974sk}:
\eq
\label{Wilson_lattice_gauge_action}
S_{\mbox{\tiny{W}}}= \beta \sum_{x \in \Lambda} \sum_{1 \le \mu < \nu \le 4} \left[1 - \frac{1}{N} \real \Tr U^{1,1}_{\mu,\nu}(x)\right],
\en
with $\beta=2N/g_0^2$ and:
\begin{equation}
U^{1,1}_{\mu,\nu}(x) = U_\mu(x) U_\nu(x+a\hat\mu) U^\dagger_\mu(x+a\hat\nu) U^\dagger_\nu(x). \label{plaquette} 
\end{equation}
However, in our study we used the tree-level improved gauge action~\cite{Weisz, Beinlich:1997ia}:
\eq
\label{tree_level_action}
S_{\mbox{\tiny{imp}}} = \beta \sum_{x\in \Lambda} \sum_{1 \le \mu < \nu \le 4} \left\{ \frac{3}{2} - \frac{1}{N}\real \Tr \left[ \frac{5}{3} U^{1,1}_{\mu,\nu}(x) - \frac{1}{12} U^{1,2}_{\mu,\nu}(x) - \frac{1}{12} U^{1,2}_{\nu,\mu}(x) \right] \right\},
\end{equation}
where:
\begin{equation}
U^{1,2}_{\mu,\nu}(x) = U_\mu(x) U_\nu(x+a\hat\mu) U_\nu(x+a\hat\mu+a\hat\nu) U^\dagger_\mu(x+2a\hat\nu) U^\dagger_\nu(x+a\hat\nu) U^\dagger_\nu(x). \label{rectangle}
\end{equation}
Assuming that the $U_\mu(x)$ group variables are related to the continuum gauge fields $A^a_\mu(x) t_a$ via: $U_\mu(x)=\exp[i a g_0 A^a_\mu(x+a\hat\mu/2) t_a]$, it is straightforward to show that both $S_{\mbox{\tiny{W}}}$ and  $S_{\mbox{\tiny{imp}}}$ tend to the Yang-Mills action in the continuum limit $a\to 0$, but the tree-level improved action defined by eq.~({\ref{tree_level_action}}) is characterized by smaller discretization effects than those of the Wilson action.
Expectation values of gauge-invariant physical observables $\mathcal{O}$ on the lattice are defined by:
\eq
\label{expectation_value}
\langle \mathcal{O} \rangle = \frac{1}{Z} \int \prod_{x \in \Lambda} \prod_{\mu=1}^4 {\dd}U_\mu(x) \; \mathcal{O} \; e^{-\Sel}
\en
and can be estimated numerically by Monte Carlo sampling over a finite set of $\{ U_\alpha(x) \}$ configurations; in the following, we denote the number of configurations used in our computations as $\nconf$. The algorithm we used to generate the configurations is based on a $3+1$ combination of local overrelaxation~\cite{overrelaxation} and heat-bath~\cite{heatbath} updates on $N(N-1)/2$ $\SU(2)$ subgroups of $\SU(N)$~\cite{Cabibbo:1982zn}. The parameters of our lattice simulations are shown in tab.~\ref{tab:simulation_info}.

\begin{table}[h]
\centering
\begin{tabular}{|c|cccccc|}  
\hline
$N$ & $N_s$ & $N_t$ & $\beta_{\mbox{\tiny{min}}}$ & $\beta_{\mbox{\tiny{max}}}$ & $n_\beta$ & $\nconf$ \\
\hline \hline
$2$ & $20$ & $5$ & $1.5$ & $16.5$ & $46$ & $2.5 \times 10^4$ \\
\hline
$3$ & $20$ & $5$ & $4$ & $7.8$ & $20$ & $1.8 \times 10^4$ \\
\hline
$4$ & $20$ & $5$ & $7$ & $7.45$ & $4$ & $2.5 \times 10^4$ \\
    & $20$ & $5$ & $7.6$ & $15.03$ & $40$ & $3 \times 10^4$ \\
    & $24$ & $5$ & $7$ & $9.85$ & $20$ & $2 \times 10^4$ \\
    & $16$ & $16$ & $7.25$ & $9.05$ & $11$ & $3 \times 10^3$ \\
\hline
$5$ & $20$ & $5$ & $12$ & $16.6$ & $30$ & $2 \times 10^4$ \\
    & $16$ & $16$ & $12.1$ & $13.7$ & $9$ & $8 \times 10^3$ \\
\hline
$6$ & $20$ & $5$ & $17$ & $25.6$ & $40$ & $2 \times 10^4$ \\
\hline
\end{tabular}
\caption{Parameters of the lattice simulations used in this work. $N$ denotes the number of colors, $N_s$ and $N_t$ are the number of sites along the space-like and time-like sizes of the lattice, $n_\beta$ is the number of $\beta$-values that were simulated, in the $\beta_{\mbox{\tiny{min}}} \le \beta \le \beta_{\mbox{\tiny{max}}}$ interval. For each set of parameters, the number of thermalized configurations, that we used in our numerical estimates, is shown in the last column.}\label{tab:simulation_info}
\end{table}

Converting the simulation results to physical units requires a definition of the lattice scale. In order to set the scale for our simulations with the improved action, we calculated the $T=0$ static potential in lattice units from expectation values of Wilson loops $\langle W(r,L) \rangle$:
\begin{equation}
\label{eq:plateau}
V(r) = a^{-1}\lim_{L\to \infty} \ln \frac{\langle W(r,L-a) \rangle}{\langle W(r,L) \rangle}.
\end{equation}
In particular, we extracted the potential from Wilson loops defined from smeared links, using five levels of smearing for the spacelike links (leaving the timelike links unsmeared). The values of $V(r)$ thus obtained are then fitted to the Cornell potential:
\begin{equation}
\label{eq:cornell}
V(r) = \sigma r + V_0 + \frac{\gamma}{r},
\end{equation}
enabling one to extract $\sigma$ (as well as $V_0$ and $\gamma$) in lattice units; statistical errors are estimated with a jackknife analysis. All fits give $\redchisq$ values close to $1$, and the $\gamma$ parameter is always very close to the bosonic string prediction: $\gamma=-\pi/12$~\cite{Luscher_string} (see fig.~1 in ref.~{\cite{Mykkanen:2011kz}}). 

Note that this non-perturbative definition of the scale is not unique: in general, it would be equally legitimate to define the value of $a$ (for a given $\beta$), using the lattice results for a different dimensionful physical observable---for example, the critical temperature $T_c$~\cite{Caselle_Panero_Piemonte}. On a finite-spacing lattice, different physical observables are generally affected by different discretization artifacts, and hence lead to slightly different definitions of the scale. This ambiguity is a systematic effect in the scale determination, but the associated relative uncertainty is numerically small, and vanishes in the continuum limit $a \to 0$. 

On the lattice, the trace of the bare Polyakov loop in the irreducible representation $r$ can be defined as:
\eq
\label{naive_lattice_Polyakov_loop}
\Tr \prod_{n_t=1}^{N_t} U^{(r)}_t(\vec{x},an_t)\;,
\en
where $g^{(r)}$ denotes the value of the group element $g$ in the irreducible representation $r$. Note that the matrix elements of a generic $g^{(r)}$ can be easily obtained from those of $g$ in the defining representation, by means of basic relations of representation theory. In particular, the characters of group elements in different irreducible representations can be easily expressed using Young calculus and the Weyl formula~\cite{Weyl_formula} (see the appendix~\ref{sec:appendix} for details). 

Note, however, that, due to the finiteness of the number of degrees of freedom on any finite lattice, the expectation value of the operator defined in eq.~(\ref{naive_lattice_Polyakov_loop}) would always be vanishing, both in the confining and in the deconfined phase. In the latter, in particular, the barriers separating different center sectors in the phase space are always finite for a finite lattice, so that any (sufficiently long, ergodic) simulation would probe all center sectors, leading to a vanishing expectation value for the average Polyakov loop. Since all numerical simulations are necessarily performed on finite lattices, it is more convenient to compute the expectation value of the \emph{modulus} of the average Polyakov loop on each gauge configuration:
\eq
\label{lattice_Polyakov_loop}
\left| \frac{1}{N_s^3} \sum_{\vec{x}} \Tr \prod_{n_t=1}^{N_t} U^{(r)}_t(\vec{x},an_t) \right| \;.
\en
Although this quantity is not an exact order parameter, it is an efficient probe of the deconfinement transition (for any irreducible representation $r$ of non-zero $N$-ality), since its expectation value tends to zero in the confining phase, while it remains finite in the deconfined phase. Henceforth, we use eq.~(\ref{lattice_Polyakov_loop}) to define the expectation values of bare Polyakov loops in our lattice simulations.

\section{Results}
\label{sec:results}

\subsection{Setting the scale}
\label{subsec:scale_subsect}

To determine the scale for our simulations with the tree-level improved lattice action, we fit our results for $a^2 \sigma$ (as extracted from Cornell fits of the $T=0$ potential using smeared Wilson loops) at the largest couplings to the functional form:
\begin{equation}
\label{scale}
a^2 \sigma = \exp \{ -[ A_0 + A_1 (\beta-\beta_0) + A_2 (\beta-\beta_0)^2 + A_3 (\beta-\beta_0)^3 ] \} ,
\end{equation}
where $\beta=2N/g_0^2$, and $\beta_0$ is an arbitrary reference value in the $\beta$-range of our simulations. 

\begin{figure*}
\centering
\vspace{-15mm}
 \includegraphics[width=.52\textwidth]{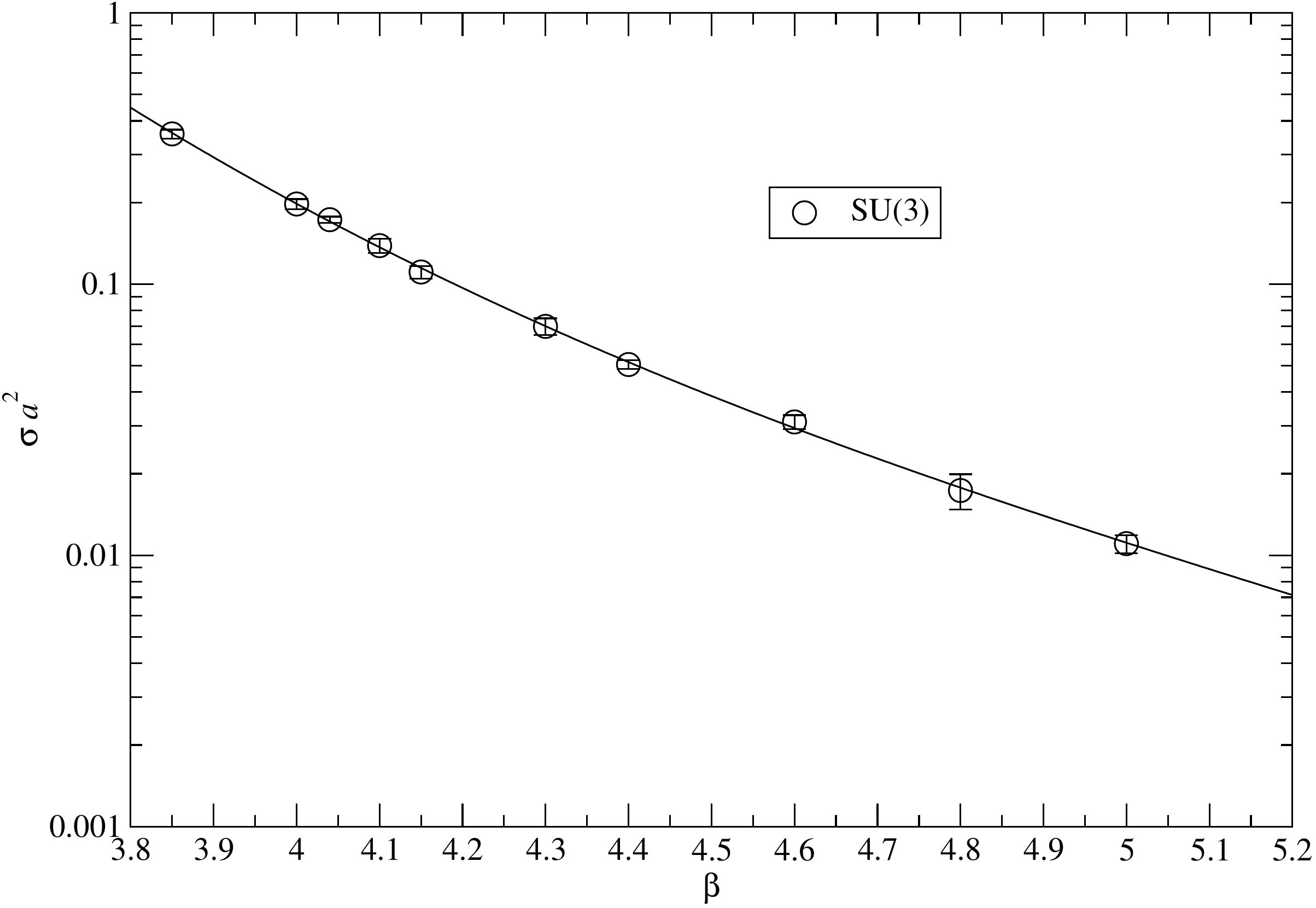}\\
\vspace{1cm}
 \includegraphics[width=.52\textwidth]{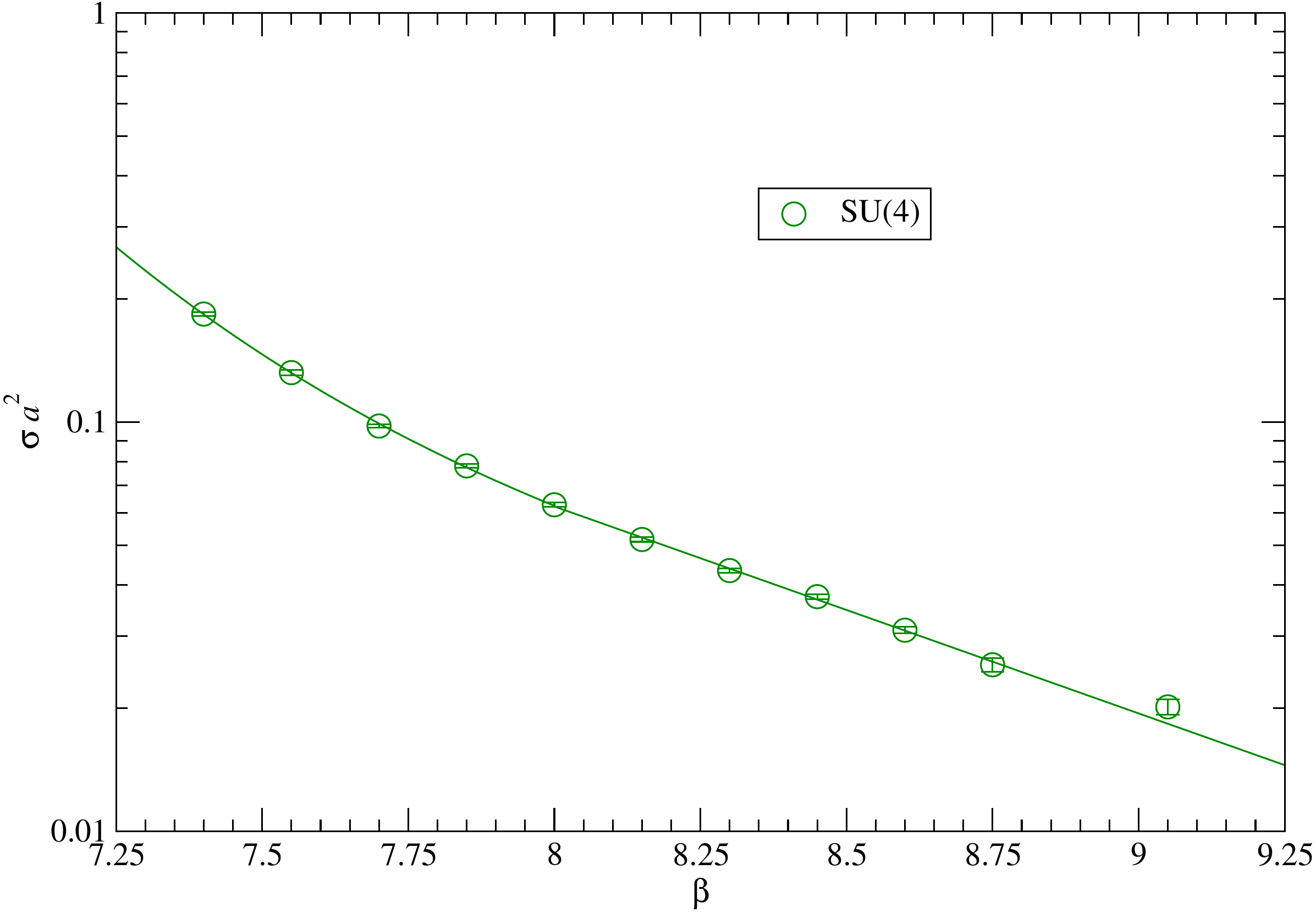}\\
\vspace{1cm}
 \includegraphics[width=.52\textwidth]{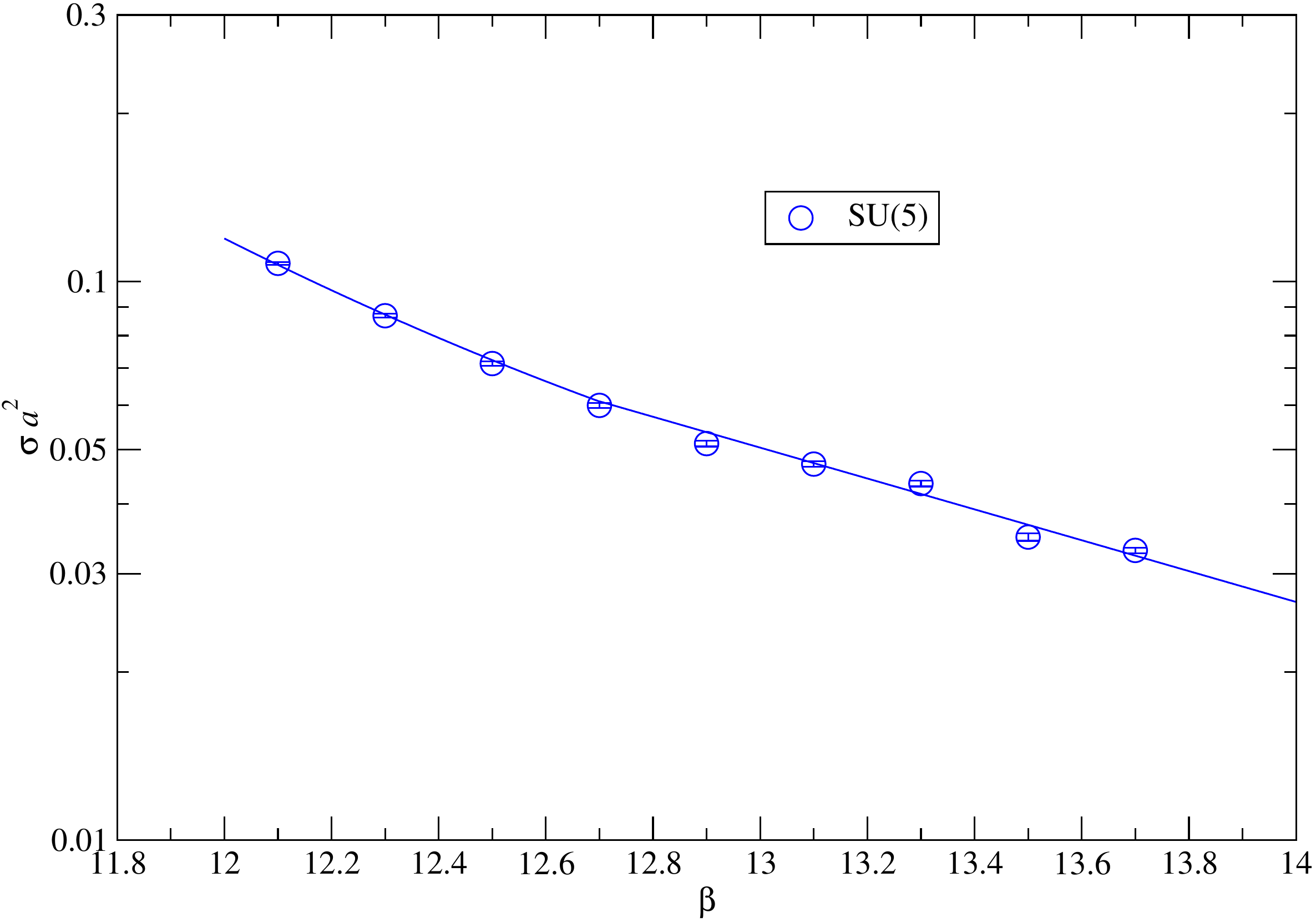}
\caption{The top panel shows a fit of the results for the string tension in lattice units in the $\SU(3)$ gauge theory, taken from ref.~\protect\cite{Beinlich:1997ia}, to the functional form in eq.~({\protect\ref{su3_scale}}). The central and bottom panels display the fits of our results for the string tension in lattice units to eq.~({\protect\ref{su4_scale}}) in $\SU(4)$ and $\SU(5)$ Yang-Mills theories, respectively.}
\label{fig:smeared_tree_level_improved_beta_string_tension_results}
\end{figure*}

As an example, fitting the $\SU(3)$ data taken from ref.~\cite{Beinlich:1997ia} to eq.~({\ref{scale}}) (choosing $\beta_0=4.3$) yields:
\begin{equation}
\label{su3_scale}
a^2 \sigma = \exp \left\{ -2.660(12) - 3.145(66)\cdot (\beta-4.3) + 0.97(11) \cdot(\beta-4.3)^2 -0.33(26)\cdot (\beta-4.3)^3  \right\},
\end{equation}
with $\redchisq=0.34$. The corresponding data, together with the fitted curve, are shown in the top panel of  fig.~\ref{fig:smeared_tree_level_improved_beta_string_tension_results}.

Similarly, our data for the $\SU(4)$ gauge group yield:
\begin{equation}
\label{su4_scale}
a^2 \sigma = \left\{ 
\begin{array}{ll}
\exp \left\{ -3.894(38) - 1.21(14) (\beta-9) - 0.41(16) (\beta-9)^2 -0.320(55) (\beta-9)^3 \right\} & \mbox{for $\beta<8$}\\ 
\exp \{ -1.165(29)\beta + 6.54(23) \} & \mbox{for $\beta \ge 8$} 
\end{array}
\right.,
\end{equation}
with $\redchisq=1.22$, and are shown in the central panel of  fig.~\ref{fig:smeared_tree_level_improved_beta_string_tension_results}, while for $\SU(5)$ we obtain:

\begin{equation}
\label{su5_scale}
a^2 \sigma = \left\{ 
\begin{array}{ll}
\exp \left\{ -3.021(15) - 0.682(17) (\beta-13) +0.214(30) (\beta-13)^2 \right\} & \mbox{for $\beta<12.7$}\\ 
\exp \{ -0.636(35)\beta + 5.28(45) \} & \mbox{for $\beta \ge 12.7$} 
\end{array}
\right.,
\end{equation}
with $\redchisq=3.41$, see the bottom panel in  fig.~\ref{fig:smeared_tree_level_improved_beta_string_tension_results}.

\subsection{Casimir scaling}
\label{subsec:casimir}

The first issue that we investigated is Casimir scaling of bare Polyakov loops, i.e., whether the free energy associated to bare Polyakov loops in a given irreducible representation $r$ is proportional to the eigenvalue of the quadratic Casimir $\langle C_2 \rangle$ of that representation. To study this problem, we rescaled the loop free energies by the ratio of the Casimir in the given representation over the one in the fundamental representation $f$. This corresponds to raising the values of the loops to the power $1/d$, where:
\begin{equation}
\label{d_definition}
d=\langle C_2 \rangle_r / \langle C_2 \rangle_f
\end{equation}
(the values of $d$ are reported in the appendix~{\ref{sec:appendix}}). 

\begin{figure*}
\centerline{\includegraphics[width=.75\textwidth]{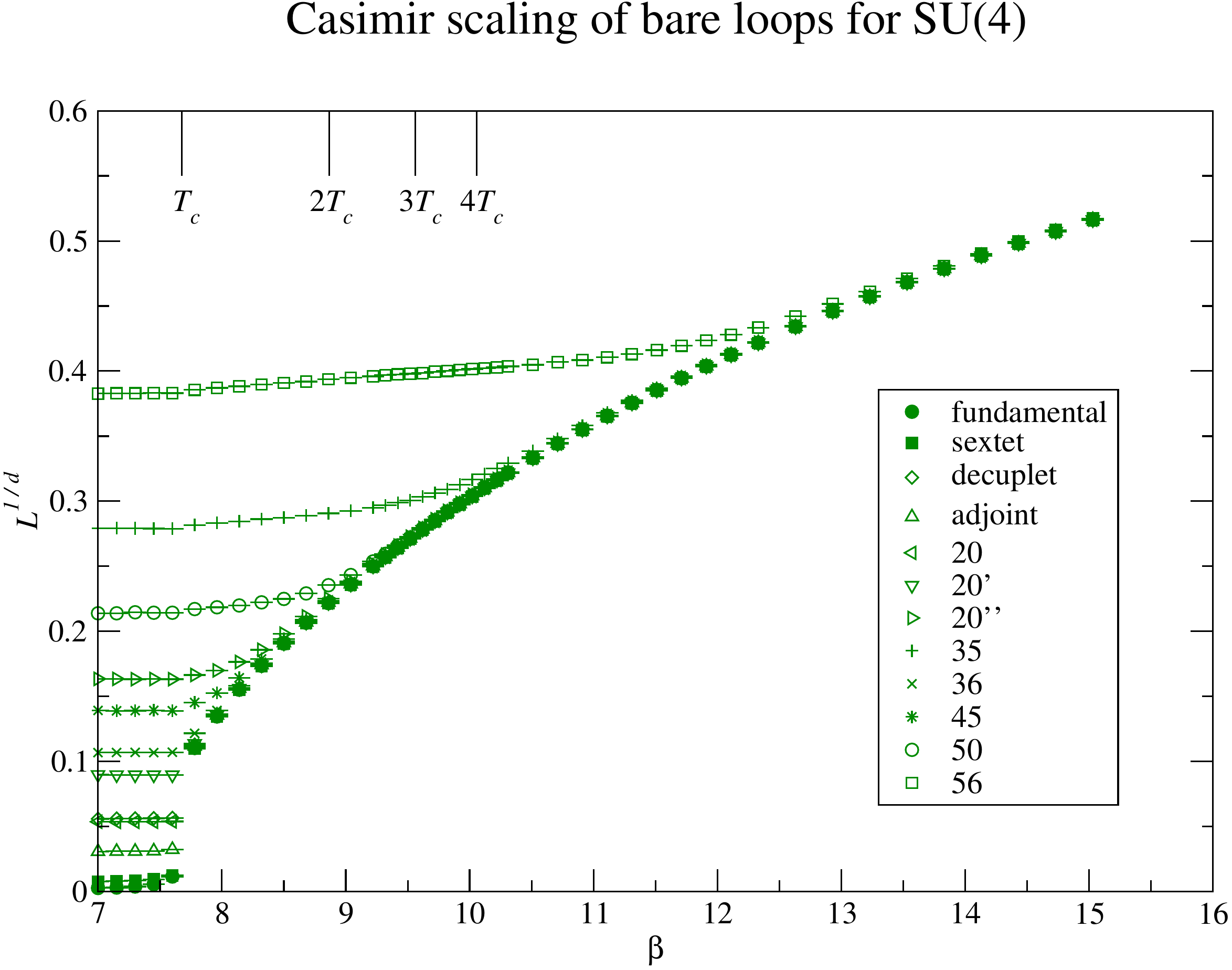}}
\caption{Temperature dependence of bare $\SU(4)$ Polyakov loops in different representations, after dividing their free energies by (a quantity proportional to) the eigenvalue of the corresponding quadratic Casimir $\langle C_2 \rangle_r$. This plot displays the results we obtained from simulations with the tree-level improved action, on lattices with $N_t=5$ and $N_s=20$ sites along the compactified time and spatial directions, respectively. The deviations from Casimir scaling observed for high representations close to the deconfinement transition are, likely, due to finite-volume effects (see the text for a detailed discussion).}
\label{fig:bare_SU4_Casimir_scaling}
\end{figure*}

Our results for the $\SU(4)$ gauge theory are displayed in fig.~\ref{fig:bare_SU4_Casimir_scaling}, which shows the values of $L^{1/d}$ for the twelve different representations, as obtained from simulations with the tree-level improved action on a lattice with $N_t=5$ and $N_s=20$ sites along the Euclidean time and spatial directions, respectively. If Casimir scaling holds, then this rescaling should make the values corresponding to higher representations collapse onto those of the fundamental representation (for which $d=1$). Note that, in this plot, the data are displayed as a function of $\beta=2N/g_0^2$: since the bare loops do not depend only on the temperature $T$, but also on the bare coupling $g_0$, it is natural to display these values (from simulations at fixed $N_t$) as a function of $\beta$. This also allows one to avoid introducing any potential ambiguity related to the definition of the temperature scale. In any case, the mapping between $\beta$ and $T$ at fixed $N_t$ is just a scale redefinition, which, for the parameters of interest, can be directly obtained combining eq.~(\ref{su4_scale}) with the relation: $T=1/(a N_t)$. In order to give an idea of the temperatures involved, we also display tick marks corresponding to a few reference temperatures along the upper horizontal axis.

Our results show an approximately perfect Casimir scaling in the deconfined phase, for all the representations that we considered. Although the bare values of loops in different representations vary by orders of magnitude, rescaling their free energies according to the corresponding quadratic Casimir eigenvalues makes them fall onto the same, universal curve. Our data show that the only significant deviations from this behavior (apart from the obvious ones in the confined phase, where Casimir scaling is \emph{not} expected to hold) are visible for strongly suppressed high representations, which are most sensitive to finite-volume effects. For example, the rescaled bare loops in the representations denoted as $\mathbf{20^{\prime\prime}}$, $\mathbf{35}$, $\mathbf{50}$ and $\mathbf{56}$ show significant deviations from the curve of the other data for temperatures $T \lesssim 1.75~T_c$, while they collapse on that curve at higher temperatures (for $L^{1/d} \gtrsim 0.2$). This is simply due to the fact that, for these representations, for $T \lesssim 1.75~T_c$ the expectation value of the corresponding loops in the thermodynamic limit is smaller than the (non-vanishing) average value of $|L|$ computed on a lattice of finite volume. This is the same effect that, on any finite lattice, is responsible for the non-vanishing values of $|L|$ in the confining phase.

\begin{figure*}
\centerline{\includegraphics[width=.45\textwidth]{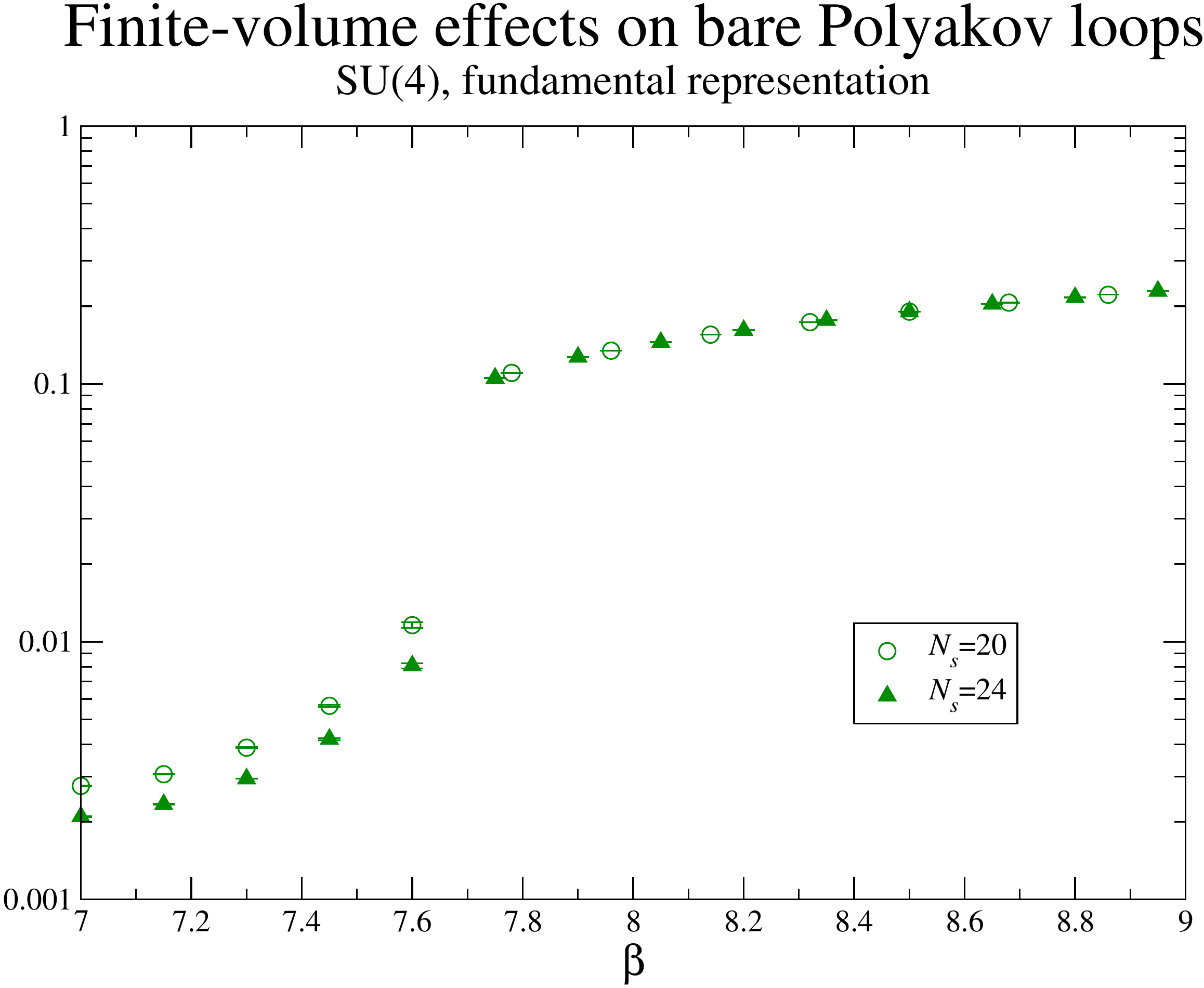} \hfill \includegraphics[width=.46\textwidth]{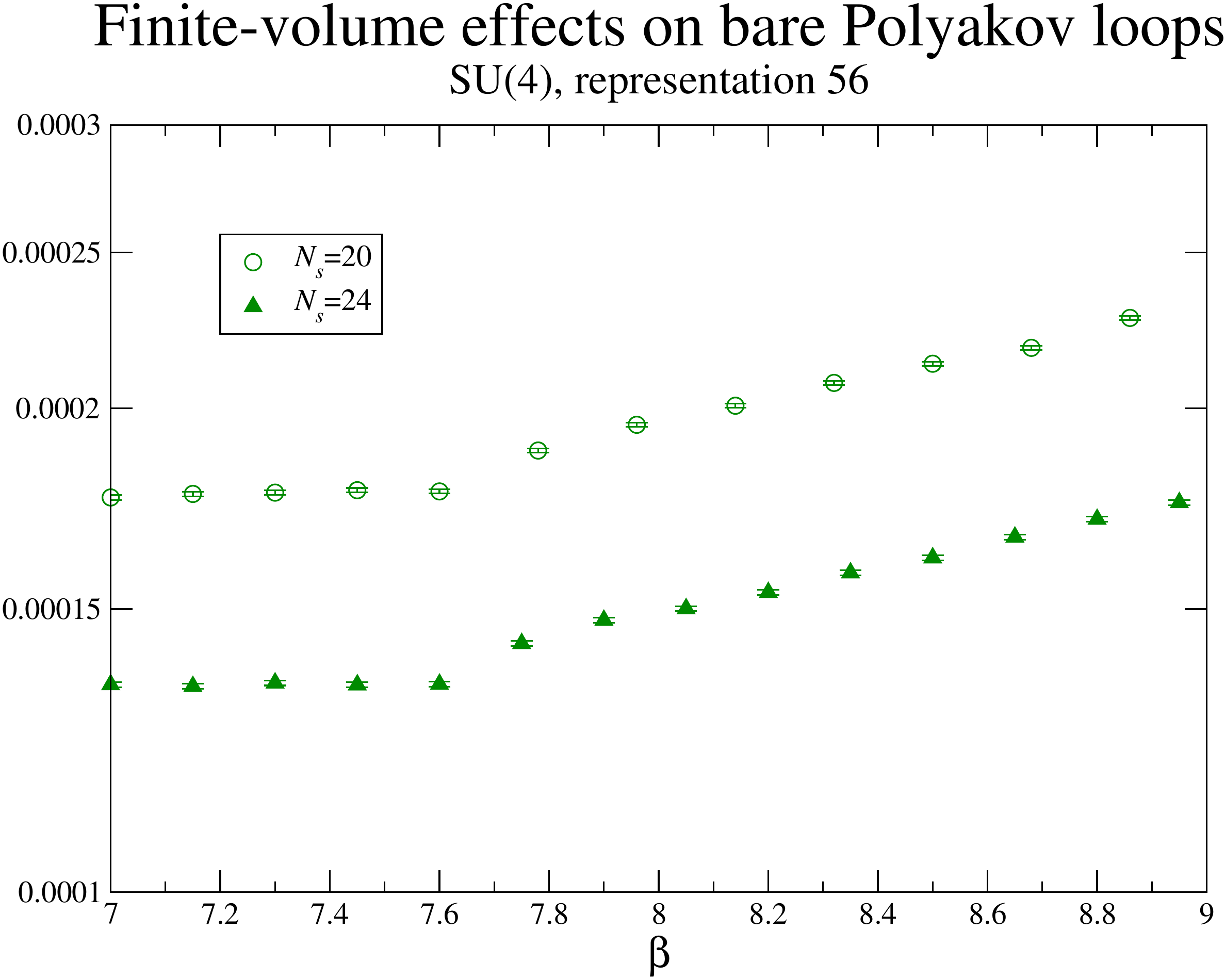}}
\caption{Left-hand side panel: Comparison of bare $\SU(4)$ Polyakov loops in the fundamental representation, obtained from lattices of different volumes: in the confined phase, the results tend to zero in the thermodynamic limit. Right-hand side panel: Loops in high-dimensional representations (such as the $\mathbf{56}$, displayed in this plot), whose expectation values are strongly suppressed, are particularly sensitive to finite-volume artifacts.}
\label{fig:finite_volume_effects}
\end{figure*}

Fig.~\ref{fig:finite_volume_effects} gives evidence of this: the left panel shows our results for bare Polyakov loops in the fundamental representation of $\SU(4)$, obtained from lattices of two different spatial volumes, $V=(20a)^3$ and $V=(24a)^3$. The results of the two sets of simulations are compatible with each other in the deconfined phase (signaling that finite-volume corrections to the critical value of $\beta$ are small for both ensembles), whereas the data obtained from the larger lattice are strongly suppressed in the confining phase, in agreement with the expectation that the average Polyakov loop is exactly zero in the thermodynamic limit. The right panel shows the same comparison, for loops in the representation of size $\mathbf{56}$: for high-dimensional representations like this, the thermodynamic limit value of the Polyakov loop is very small, even in large regions of the deconfined phase, and thus it is overwhelmed by finite-size artifacts on the lattices that we considered. As fig.~\ref{fig:bare_SU4_Casimir_scaling} shows, for such representations it is only at very large values of $\beta$ (i.e., at very high temperatures) that the contribution surviving the thermodynamic limit becomes dominant over finite-volume artifacts. 

In principle, one could perform an extrapolation to the thermodynamic limit, by repeating the simulations on a series of lattices of increasing volume. However, it should be pointed out that this would require a non-trivial computational effort for higher representations, especially at temperatures close to the deconfinement region. While this task is beyond the scope of the present work, we emphasize that the results displayed in the right panel of fig.~\ref{fig:finite_volume_effects} give strong support to the interpretation of the deviation from Casimir scaling for high representations close to the deconfinement region in our data as a phenomenon which is (at least partially) due finite-volume artifacts. In particular, this plot (in which the scale on the vertical axis is logarithmic) shows that, for this high representation, an increase of the lattice volume by a factor approximately equal to $1.73$ leads to a nearly uniform shift of all data towards smaller values, and that this happens both in the confined and in the deconfined phase. The comparison with the left panel, which shows that in the same range of couplings (i.e., of temperatures) and for the same values of $V$, our numerical results for the fundamental representation are sensitive to this shift only in the confined phase, is strongly suggestive that, at temperatures close to $T_c$, the numerical data for high representations are dominated by finite-volume effects, and, hence, that the deviations from Casimir scaling observed in fig.~\ref{fig:bare_SU4_Casimir_scaling} do not necessarily survive in the thermodynamic limit.

\begin{figure*}
\centerline{\includegraphics[width=.48\textwidth]{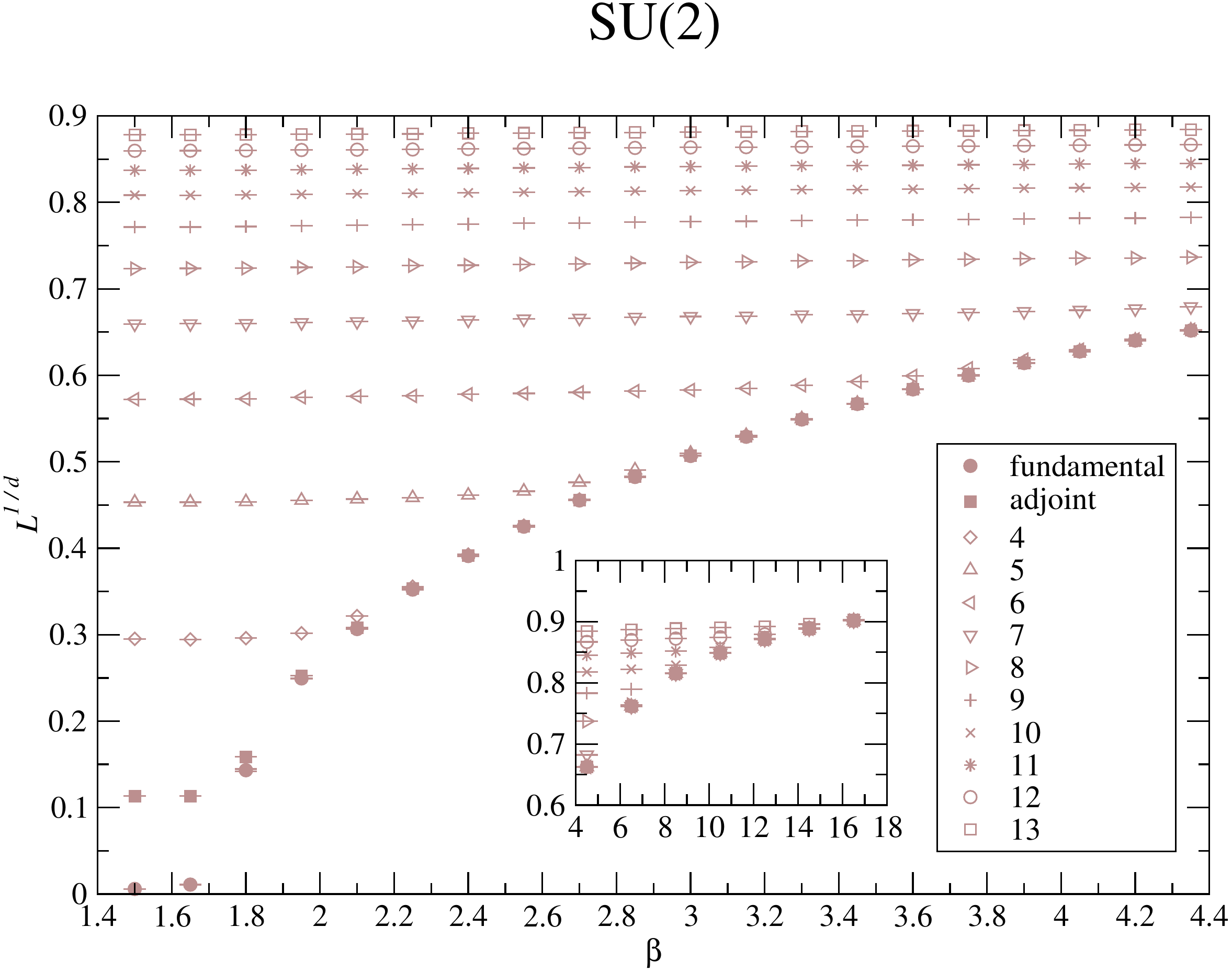} \hfill \includegraphics[width=.48\textwidth]{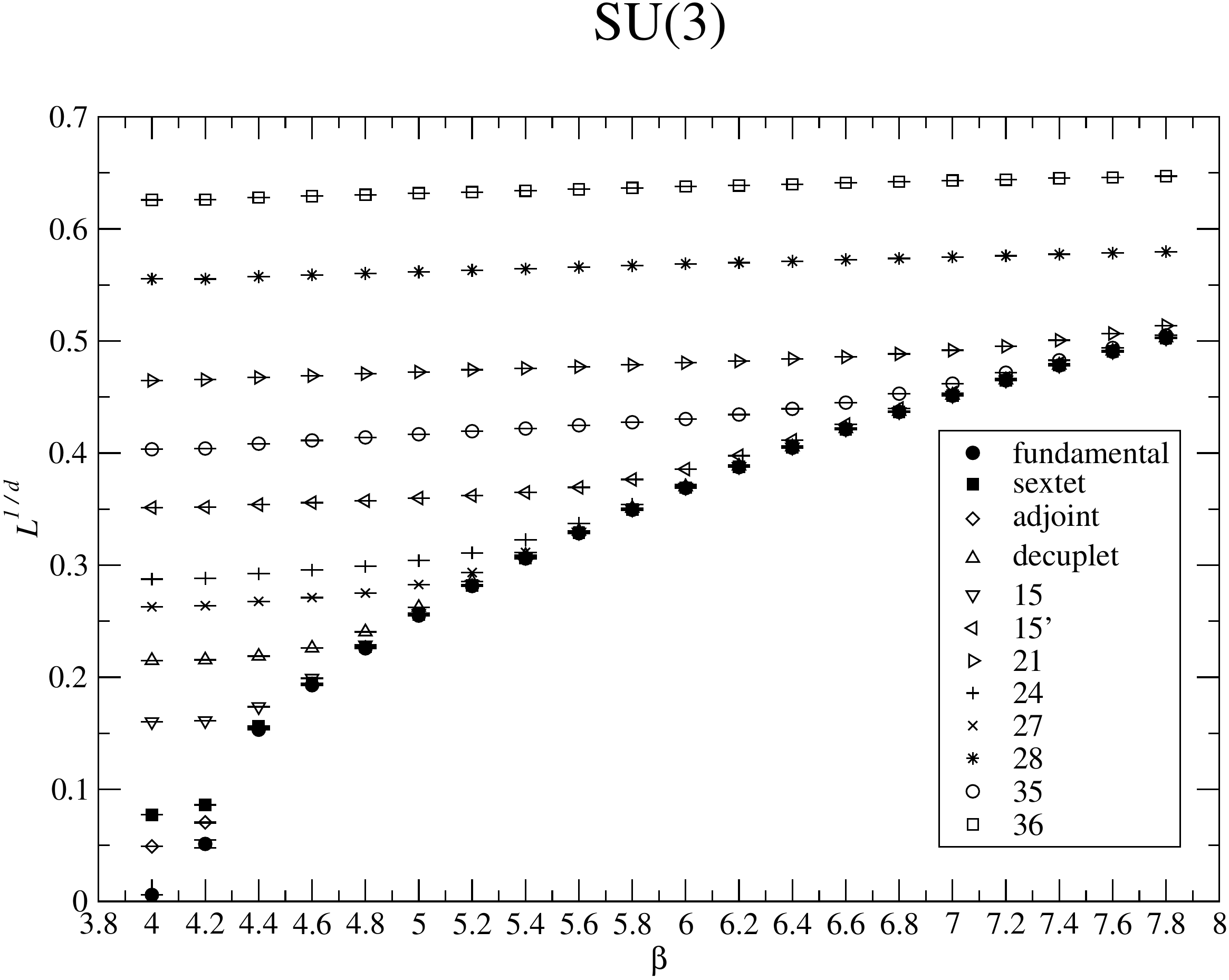}}
\vspace{5mm}
\centerline{\includegraphics[width=.48\textwidth]{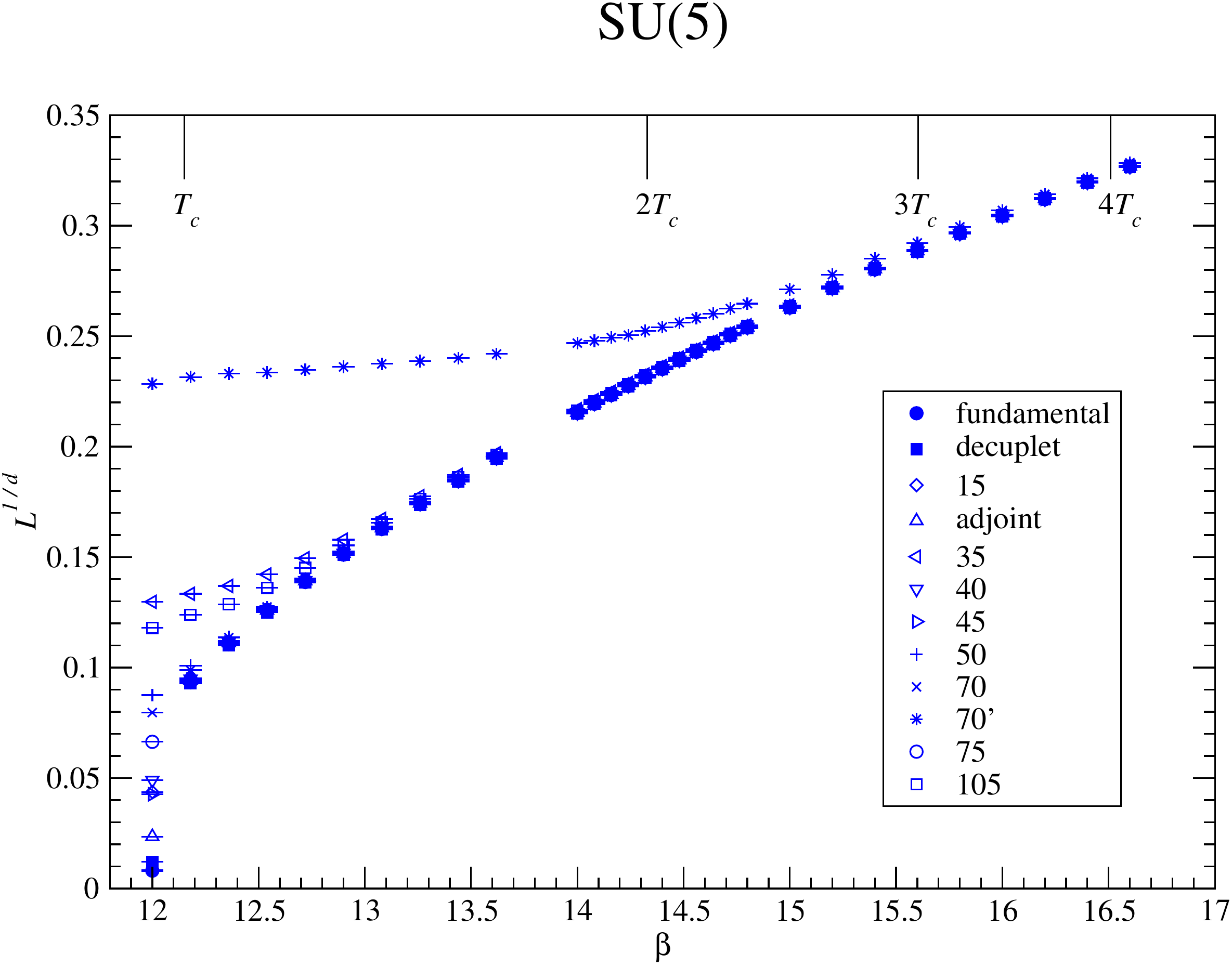} \hfill \includegraphics[width=.48\textwidth]{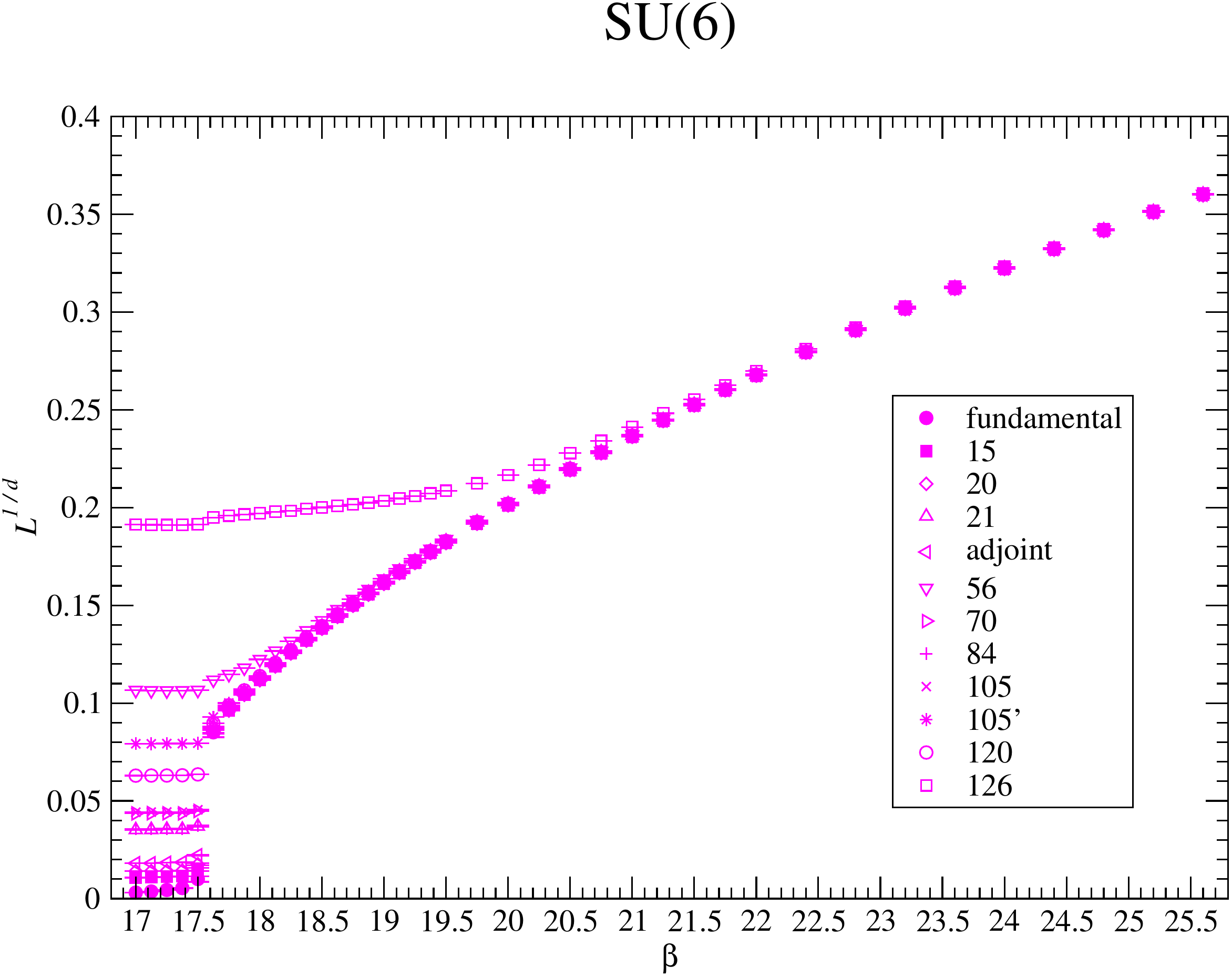}}
\caption{The top left panel shows results analogous to those in fig.~\protect\ref{fig:bare_SU4_Casimir_scaling}, but for the $\SU(2)$ Yang-Mills theory. The inset shows the convergence to a universal curve, in a parameter range where finite-volume effects cease to dominate the results for higher representations. Similarly, the top right panel shows the corresponding results for the $\SU(3)$ gauge group, whereas the bottom left and bottom right plots display the results for the $\SU(5)$ and $\SU(6)$ theories, respectively.}
\label{fig:bare_othergroups_Casimir_scaling}
\end{figure*}

Our results for bare Polyakov loops in different representations (rescaled by dividing the respective free energies by the factor $d$, proportional to the quadratic Casimir eigenvalue of the corresponding representation) for the $\SU(2)$, $\SU(3)$, $\SU(5)$ and $\SU(6)$ theories are displayed in fig.~\ref{fig:bare_othergroups_Casimir_scaling}: they reveal the same behavior observed for the $\SU(4)$ gauge group. Furthermore, comparing the plots of the rescaled bare loops for different groups, one also observes that, when $N$ grows, the numerical study of higher representations simplifies, in the sense that they tend to be less sensitive to finite-volume effects. This is related to the fact that, in general, for $N \to \infty$ the quadratic Casimir grows only linearly with $N$, and with the number of fundamental and anti-fundamental indices out of which a generic representation is built (see the appendix~\ref{sec:appendix} for a discussion). For example, while the $d$ factor for the highest $\SU(2)$ representation considered here (i.e., for the twelfth lowest, non-trivial) is equal to $56$, its value for the twelfth $\SU(6)$ representation is less than $6$. As a consequence, from this point of view, the study of higher representations at large $N$ actually becomes \emph{simpler} than for smaller gauge groups.

Note that, deep in the weak-coupling region, one could compare the simulation results with the predictions from lattice perturbation theory. In particular, for the Wilson action the latter have been known for many years~\cite{Heller:1984hx}. However, since our simulations are based on the tree-level improved action, rather than the Wilson action, we did not perform such a comparison.

\subsection{Renormalized Polyakov loops}
\label{subsec:renormalized}

Finally, we present our results for the renormalized Polyakov loops, restricting our attention to loops in the fundamental representation. Our renormalization procedure is based on the determination of the constant term $V_0$ in the $T=0$ interquark potential extracted from the lattice, at each value of the bare coupling. More precisely, we define the renormalized Polyakov loop as:
\begin{equation}
\label{our_renormalization_prescription}
\langle \Lren \rangle = Z^{N_t} \langle L \rangle, \qquad \mbox{with: } Z=e^{a V_0/2},
\end{equation}
where $aV_0$ is obtained from our fits of the interquark potential. Note that, in the expression above, the renormalized loop $\Lren$ is expected to depend only on the physical temperature $T$, while the bare one $L$ depends both on $g_0^2$ and on $N_t$. On the other hand, $Z$ depends only on $g_0^2$. 

Note that, since eq.~(\ref{our_renormalization_prescription}) defines $ \langle \Lren \rangle $ in the fundamental representation through the charge renormalization factor $Z$, it follows that the corresponding factors for any higher representation can be defined as $Z^d$, with $d$ defined in eq.~(\ref{d_definition}).\footnote{One could also imagine to define the renormalization factor for a higher representation $r$, by extracting the constant term of the potential between static sources in that representation. However, this procedure would be very tricky, for several reasons. In particular, sources in representations of vanishing $N$-ality at $T=0$ get completely screened at large distances, while for representations of non-zero $N$-ality it is well-known that the confining behavior at asymptotically large distances is characterized by the string tension of the smallest representation with the same $N$-ality (although, at intermediate distances, the slope of the confining potential can be different). Moreover, extracting the confining potential from lattice calculations of Wilson loops in higher representations is computationally very demanding, due to the strong suppression of the signal-to-noise ratio, which is exponentially damped with the loop sizes and with the string tension. These features make a proper definition of $a V_0$ for high representations subtle, and its extraction from lattice simulations particularly challenging.} As a consequence, with this renormalization procedure, it follows that renormalized Polyakov loops in higher representations obey Casimir scaling, if the corresponding bare ones do. This is no longer necessarily true, if a different renormalization prescription is used (see below for a discussion). However, previous studies of the $\SU(3)$ gauge theory revealed that renormalized loops still obey Casimir scaling to very high accuracy, even when different renormalization methods (involving renormalization factors which are, a priori, independent for each representation) are used---see, e.g., ref.~\cite{Gupta:2007ax}. Since these alternative renormalization methods are, typically, quite noisy and not ideally suited for computationally demanding simulations of $\SU(N>3)$ gauge theories, in the present work we restricted our analysis to the renormalization prescription defined by eq.~(\ref{our_renormalization_prescription}), focusing our attention on the fundamental representation.

\begin{figure*}
\centerline{\includegraphics[width=.7\textwidth]{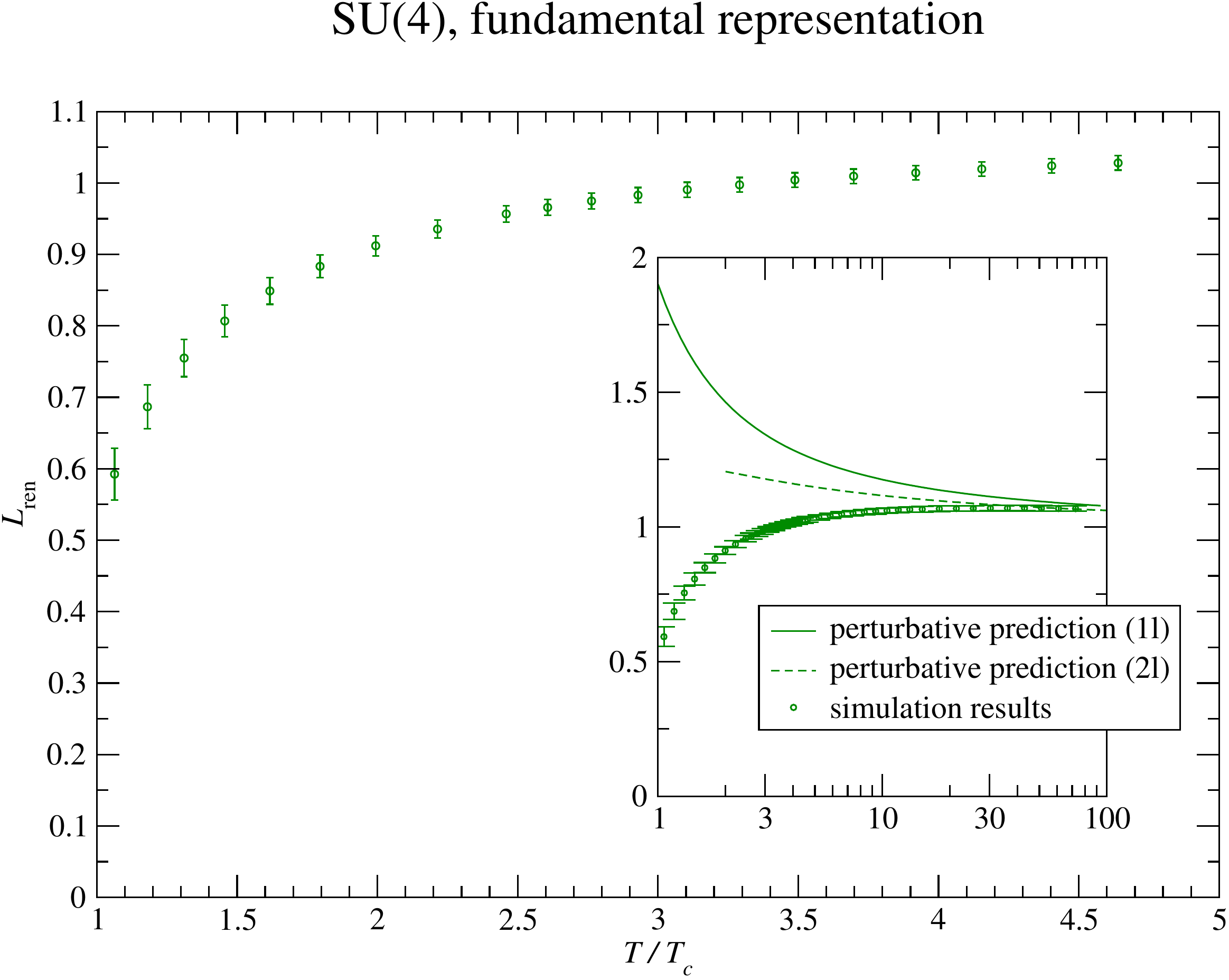}}
\vspace{5mm}
\centerline{\includegraphics[width=.7\textwidth]{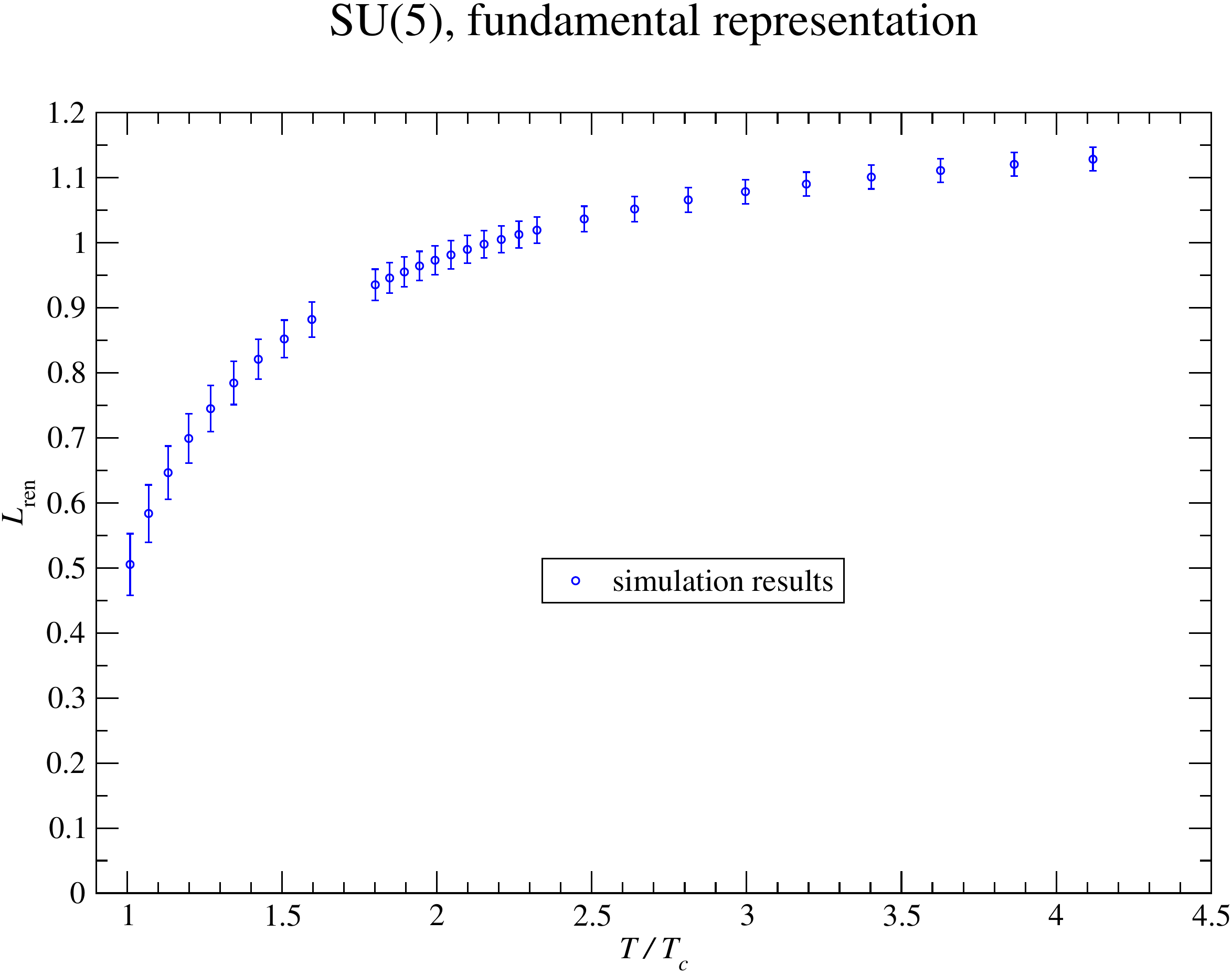}}
\caption{Top panel: Renormalized $\SU(4)$ Polyakov loop in the fundamental representation, as a function of the temperature (in units of $T_c$), in comparison with one- and two-loop  perturbative predictions. Bottom panel: Renormalized fundamental loop, as a function of $T/T_c$, in the $\SU(5)$ theory.}
\label{fig:renormalized_loop}
\end{figure*}

For $\SU(4)$, in the temperature region that we are most interested in (i.e., in the deconfined phase, close to $T_c$), our fits show that an accurate parameterization of $Z(g_0^2)$ is of the form:
\begin{equation}
\label{Z_su4}
Z(g_0^2)=\exp[-0.166(21)g_0^2 +0.259(28)g_0^4],
\end{equation}
for $g_0^2 \le 0.8$; the quoted errors are conservative. Using eq.~({\ref{Z_su4}}) to renormalize the bare loops obtained from our simulations with $N_t=5$, $N_s=20$, we obtain the renormalized Polyakov loop values displayed in the top panel of  fig.~\ref{fig:renormalized_loop}, in which the displayed errorbars also include an estimate of the systematic uncertainties related to scale setting and renormalization prescription choice (which are discussed below). The inset shows a comparison of our data over a broader range (with extrapolation in the scale setting and in the parameterization of the renormalization constant) with the perturbative prediction for this gauge group, taken from ref.~\cite{perturbative_Polyakov_loop}. In particular, the upper solid curve is obtained using one-loop estimates for the coupling and Debye mass, whereas the lower dashed curve is obtained from two-loop estimates of these quantities~\cite{twoloopestimates}. The figure shows that the renormalized loop takes a value close to $1/2$ for $T \to T_c^+$, and increases with the temperature, overshooting $1$ at $T\simeq 3.4 T_c$. Extrapolating our parameterizations for the scale and for $Z$ to a range of small coupling values (in which we have not performed non-perturbative computations of the $T=0$ interquark potential), we find that the renormalized fundamental loop in the $\SU(4)$ theory reaches a maximum (about $1.07$) at temperatures around $30 T_c$, then starts decreasing and approaching the next-to-next-to-leading order perturbative prediction, which, for the $\SU(N)$ Yang-Mills theory, reads~\cite{perturbative_Polyakov_loop}:
\begin{equation}
\label{perturbative_prediction}
\Lren = 1 + \frac{g^2 m_E \langle C_2 \rangle_f }{8\pi T} + \frac{g^4 N \langle C_2 \rangle_f }{(4\pi)^2} \left( \ln \frac{m_E}{T} + \frac{1}{4} \right) + O(g^5),
\end{equation}
where $g$ denotes the physical coupling, and $m_E$ is the Debye mass. The behavior we observe in our $\SU(4)$ data is consistent with the results obtained for $\SU(3)$ in previous studies~\cite{Kaczmarek:2002mc, Dumitru:2003hp, Gupta:2007ax}.

Similarly, our results for the $\SU(5)$ gauge group are based on the following parameterization of $Z(g_0^2)$:
\begin{equation}
\label{Z_su5}
Z(g_0^2)=\exp[0.4115(26)g_0^2],
\end{equation}
for $g_0^2 \le 0.8$, and are displayed in the bottom panel of fig.~\ref{fig:renormalized_loop}. Similarly to the case of four colors, also in the $\SU(5)$ theory the renormalized loop has a value close to $1/2$ for $T \to T_c^+$, and increases up to values larger than $1$.

\begin{figure*}
\centerline{\includegraphics[width=.45\textwidth]{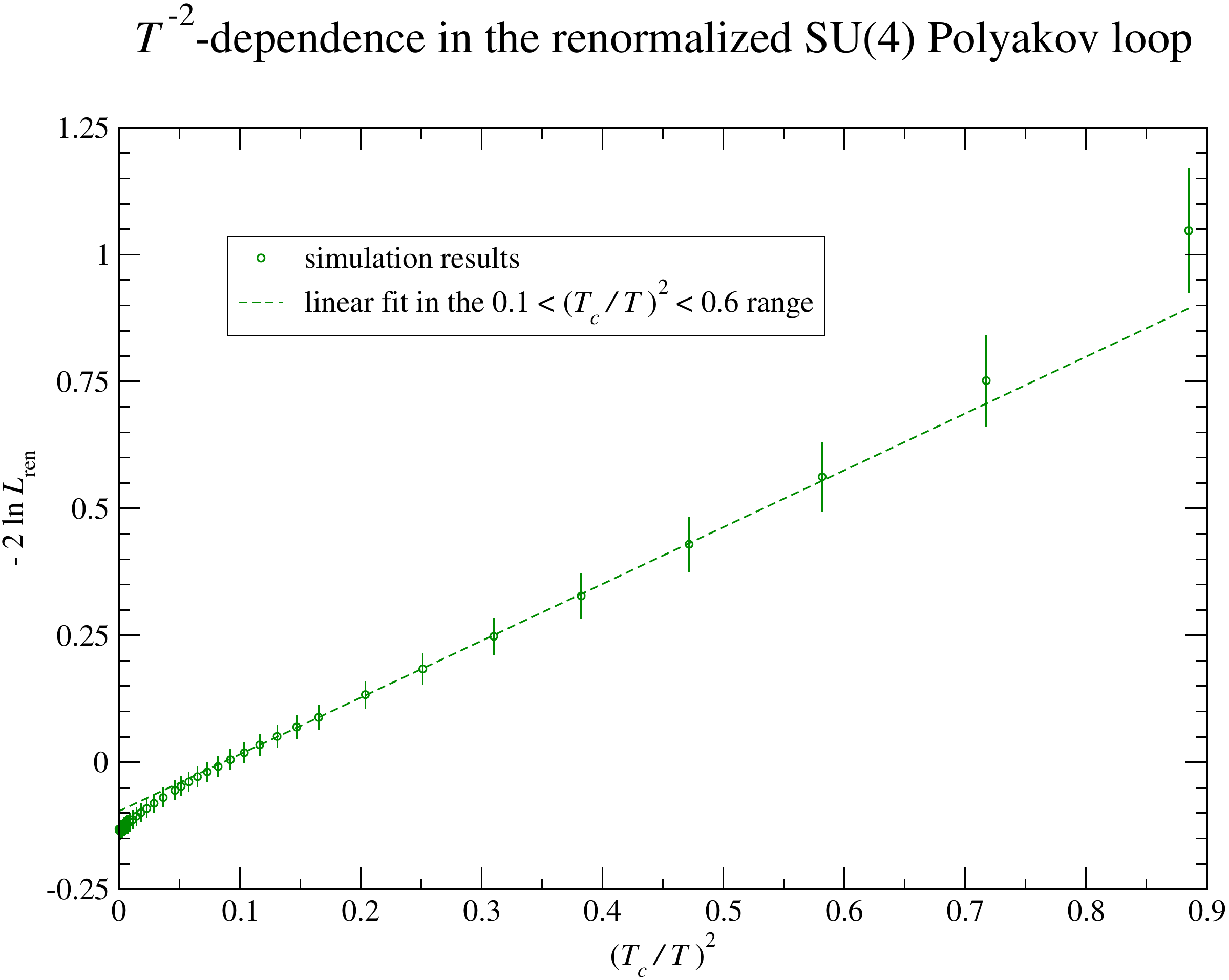} \hfill \includegraphics[width=.45\textwidth]{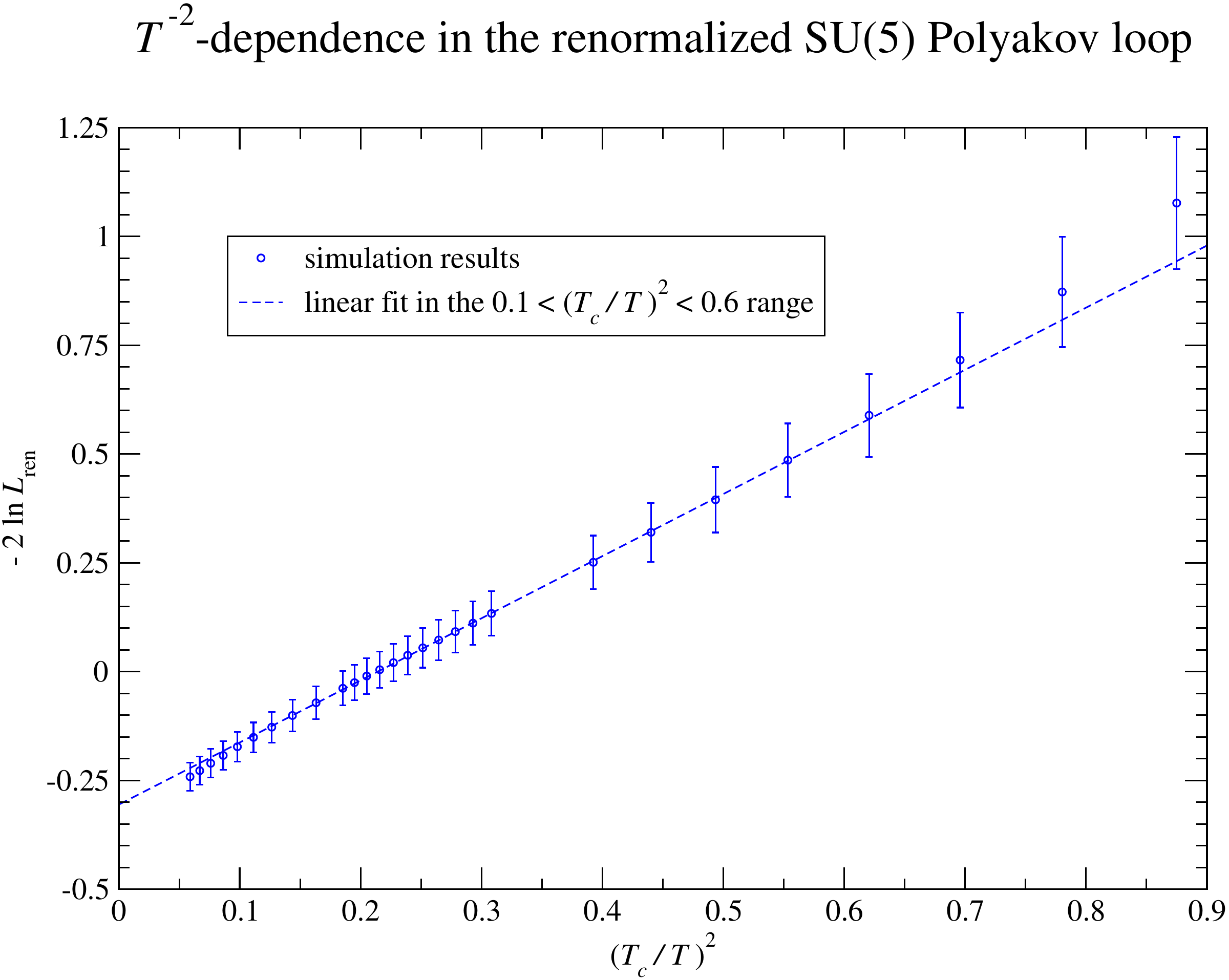}}
\caption{Similarly to what was observed for the $\SU(3)$ theory~\cite{Megias_Ruiz_Arriola_Salcedo}, also the logarithm of the $\SU(4)$ (left-hand side panel) and $\SU(5)$ (right-hand side panel) renormalized Polyakov loops exhibits a characteristic $T^{-2}$-dependence in the deconfined phase, up to temperatures of a few times $T_c$. Note that the errorbars include conservative estimates of the systematic uncertainties (see subsection~{\protect\ref{subsec:systematics}}).}
\label{fig:Tsquare}
\end{figure*}

Finally, in fig.~\ref{fig:Tsquare}, we show (minus twice) the logarithm of the renormalized fundamental loops for the $\SU(4)$ and $\SU(5)$ gauge groups, as a function of the inverse of the square of the temperature. As it was already observed in the case of the $\SU(3)$ theory~\cite{Megias_Ruiz_Arriola_Salcedo}, at temperatures between $T_c$ and a few times $T_c$, the logarithm of the renormalized Polyakov loop appears to be of the form:
\begin{equation}
\label{constant_plus_one_over_Tsquare}
-2 \ln \Lren = m \left(\frac{T_c}{T}\right)^2 + q.
\end{equation}
In fig.~\ref{fig:Tsquare}, the straight lines are fits (in the temperature ranges shown in the plot legends) to eq.~({\ref{constant_plus_one_over_Tsquare}}), which yield $m=1.1166(55)$, $q=-0.0959(11)$, with $\redchisq = 0.004$ for $\SU(4)$ and $m=1.4283(62)$, $q=-0.3056(15)$, with $\redchisq = 0.003$ for $\SU(5)$. The small $\redchisq$ values are due to the fact that the errorbars affecting our numerical results are dominated by the systematic uncertainties, for which we could only provide a crude, but conservative, estimate.

Note, however, that the statement, that the logarithm of the renormalized Polyakov loop is of the form appearing on the right-hand side of eq.~({\ref{constant_plus_one_over_Tsquare}}), is a scheme-dependent one. For example, redefining the renormalized Polyakov loop free energy with the addition of a constant, would introduce an additive contribution $O(T^{-1})$ to the logarithm of the renormalized loop. We find that the statement holds for the renormalization scheme that we discussed here (see also ref.~{\cite{Kaczmarek:2002mc}}). 

In view of this observation, one may wonder, whether there are arguments supporting our scheme choice, rather than others. As discussed above, our renormalization scheme for the Polyakov loop is based on the subtraction of the constant term appearing in the $T=0$ potential between two static sources. This reduces the form of the renormalized confining potential to:
\begin{equation}
V(r) = \sigma r + \frac{\gamma}{r} + O(r^{-2}).
\label{stringpotential}
\end{equation}
The functional form appearing on the right-hand side of eq.~({\ref{stringpotential}}) can be derived (at the leading order in an expansion around the large-distance limit) from an effective bosonic string model for confinement~\cite{Luscher_string}. Various recent works (see, e.g., ref.~\cite{effective_string} and references therein) show that Lorentz-Poincar\'e symmetries constrain the first few terms in the expansion of the effective string action to equal those that are obtained expanding the Nambu--Goto action~\cite{Nambu_Goto}, while corrections only appear at high orders in $1/r$. The fact that the Nambu--Goto string provides a good effective model for the confining potential is also confirmed by extensive numerical evidence from lattice simulations, both for $\SU(N)$ Yang-Mills theories~\cite{Teper:2009uf} and for theories based on smaller gauge groups~\cite{string_in_small_groups}. Following ref.~\cite{Hidaka:2009xh}, it is then natural to define a renormalization scheme yielding a $T=0$ interquark potential with a vanishing constant term, eq.~({\ref{stringpotential}}), and to apply it to the renormalization of the Polyakov loop.

In ref.~\cite{Andreev:2009zk}, a holographic prediction for the renormalized Polyakov loop was computed, using a model with one deformation parameter~\cite{Andreev:2007zv}. The result reads:
\begin{equation}
\label{andreev_prediction}
\Lren (T) = b_1 \exp \left\{ -b_2 \left[ \sqrt{\pi} \frac{T_c}{T}\mbox{Erfi}\left( \frac{T_c}{T} \right) - \exp \left( \frac{T_c}{T}\right)^2 \right] \right\},
\end{equation}
where $b_1$ and $b_2$ are two coefficients that can be fitted, and $\mbox{Erfi}$ denotes the imaginary error function. At the leading order in a high-temperature expansion, eq.~({\ref{andreev_prediction}}) predicts that the logarithm of the  Polyakov loop would be given by the sum of a constant plus a $(T_c/T)^2$ term, as observed in the numerical data. 

More recently, a holographic computation of the Polyakov loop was also performed in ref.~\cite{Megias:2010ku}, finding good agreement with the $\SU(3)$ lattice data from ref.~\cite{Gupta:2007ax}, and a numerical value of $\Lren(T)$ very close to $1/2$ for $T \to T_c^+$.

In the literature, it was suggested that the dependence of $\ln \Lren$ on $T^{-2}$ could be due to a non-perturbative contribution from a gluon condensate~\cite{Megias_Ruiz_Arriola_Salcedo, Xu:2011ud}. Similar arguments have been invoked to explain the behavior of the interaction measure $\Delta$ at temperatures of the order of a few times $T_c$~\cite{Meisinger:2001cq, Megias:2005ve, Tsquare_in_Delta, Andreev:2007zv}: in all $\SU(N)$ gauge theories, both in $D=3+1$~\cite{largeN3+1D} and in $D=2+1$~\cite{largeN2+1D} spacetime dimensions, $\Delta$ appears to be proportional to $T^2$.

\subsection{Systematic uncertainties}
\label{subsec:systematics}

Apart from the precision limits related to the finiteness of our statistical samples, the main systematic uncertainties affecting our study include: ambiguities in the scale determination, renormalization prescription dependence, effects due to the volume finiteness, and finite-cutoff effects. Let us discuss each of them in turn.

In the temperature range of interest in this study, a reliable definition of the temperature scale is necessarily non-perturbative, and---as discussed above---requires the choice of a dimensionful physical observable of reference. As different observables are generally affected by different lattice artifacts, this leads to slight ambiguities in the definition of the scale; however, this systematic effect becomes negligible at small lattice spacings. A potentially more severe ambiguity is related to the functional form that one can choose to parameterize the data to be fitted. Rather than interpolating our simulation results with arbitrary, arbitrarily complicated functions, in the present study we tried to use physically motivated functional forms, with a minimal number of parameters, and estimated the systematic uncertainty related to scale setting by comparing the results with different parameterizations, at various values of the lattice gauge coupling.

A potentially large systematic ambiguity in our computation is related to the choice of the Polyakov loop renormalization method. In the present work, we followed the approach already used in a similar study for the $\SU(2)$ gauge theory~\cite{Hubner:2008ef}. Other related studies discuss different renormalization methods, which lead to roughly compatible results. In particular, the authors of ref.~\cite{Gupta:2007ax} discussed a comparison of a renormalization method based on the $Q\bar{Q}$ potential (similar to our prescription) with an iterative renormalization (based on simulations on lattices of different spacing and at the same temperatures): these two methods appear to be compatible with each other, although the latter has the drawback of leading to an accumulation of statistical errors, particularly at temperatures close to $T_c$. A different renormalization method was suggested in ref.~\cite{Dumitru:2003hp}: there, the idea is to extract the free energy of the renormalized Polyakov loop at a given temperature $T$, by identifying the $N_t$-independent contributions to the free energy $F$ of bare loops extracted from simulations on lattices of different spacing:
\begin{equation}
F = N_t F^{\mbox{\tiny{div}}} + \Fren + F^{\mbox{\tiny{lat}}}/N_t + \dots,
\end{equation}
where $F^{\mbox{\tiny{div}}}$ and $F^{\mbox{\tiny{lat}}}$ respectively denote the coefficient of the contribution to the bare free energy that diverges in the continuum limit, and the coefficient of the leading term due to lattice artifacts. 
A problem with this method, however, is that, in order to keep the temperature $T=1/(aN_t)$ fixed, the lattice spacing $a$ is obviously different for each simulation at a different $N_t$. Since $a$ is tuned by varying the bare coupling $g_0$, this implies that $F^{\mbox{\tiny{div}}}$, $\Fren$ and $F^{\mbox{\tiny{lat}}}$, which generically depend on $g_0$, are \emph{not} held fixed when $T$ is fixed. Yet another renormalization method was proposed in ref.~\cite{Gavai:2010qd}, following the fixed-scale approach~\cite{Umeda:2008bd}: the idea is to fix $Z$ at only one value of the bare coupling $g_0$, and then to vary the temperature in the lattice simulations by varying $N_t$ at fixed spacing $a$ (i.e., at fixed bare coupling). A potential drawback of this method, however, is that it does not allow one to vary the temperature continuously. Aspects related to the renormalization of Wilson lines have also been discussed in ref.~\cite{Hidaka:2009xh}. To get a rough estimate of the systematic uncertainty associated with the choice of a renormalization method, we compared the difference between various methods, at different temperatures, both in our data and in the results available in the literature.

Finally, finite-volume and finite-cutoff effects appear to be under control in our study. In particular, the results of our simulations show that, for $N \ge 4$, deviations from the thermodynamic limit in the deconfined phase are clearly visible only for high representations, whereas they appear to be negligible for the fundamental representation (see fig.~{\ref{fig:finite_volume_effects}}). In fact, the lattices used in the present study are characterized by an aspect ratio $N_s/N_t \ge 4$, which is known to provide a good approximation of the thermodynamic limit in the temperature range of interest~\cite{finite_volume_finite_T}. As for finite-cutoff effects, unfortunately we could not repeat all of our calculations on finer lattices, hence we are unable to perform a continuum extrapolation of our results. The systematic error due to cutoff effects on lattices with $N_t=5$, however, is expected to be rather small for simulations with the improved action that we used: previous studies for the $\SU(3)$ gauge group~\cite{Gupta:2007ax} showed no significant discrepancies between $N_t=4$ and $N_t=8$.

Adding up the various sources of systematic uncertainties in quadrature, the total relative errors on our renormalized Polyakov loops are in the range between $1\%$ and $5\%$ for $\SU(4)$, and between $1.5\%$ and $8\%$ for $\SU(5)$.

\section{Discussion and conclusions}
\label{sec:conclusions}

Our main findings can be summarized as follows.
\begin{enumerate}
\item For all gauge groups, and for all the representations considered in this work, the bare Polyakov loops show excellent Casimir scaling, for all values of the coupling (or, equivalently, of the temperature), down to the deconfining transition. The only deviations, that our data reveal, can be explained in terms of finite-volume artifacts: they especially affect the high representations, particularly close to $T_c$ (while they become negligible at sufficiently high temperatures), and can be reduced by increasing the spatial volume of the system. As we mentioned, in the literature, several works have reported evidence for Casimir scaling in $\SU(N)$ Yang-Mills theories, including, in particular, for the $T=0$ string tension associated to the potential between two static sources in a given representation~\cite{Casimir_scaling_references}. This observation has an interesting implication related to the large-$N$ limit. As it is well-known, for pure $\SU(N)$ Yang-Mills theories, the expansion around the $N \to \infty$ limit can be organized in a series of powers of $1/N^2$, i.e. it does not contain odd powers of $1/N$. As discussed in ref.~\cite{Armoni_Shifman}, this expectation seems to be at odds with the numerical evidence of Casimir scaling of $k$-string tension from lattice simulations~\cite{Casimir_scaling_references}, since, in general, if Casimir scaling holds, then the leading finite-$N$ corrections are $O(1/N)$. The resolution of this apparent paradox, however, was recently pointed out in ref.~\cite{Greensite:2011gg}, and is based on a cancellation of terms involving odd $1/N$ powers in the spectrum of open string states. Similar arguments were also discussed in ref.~\cite{KorthalsAltes:2005ph}.
\item In the thermodynamic limit the renormalized fundamental Polyakov loop is vanishing in the confined phase, and jumps to a finite value at the critical temperature, compatibly with the first-order nature of the deconfining transition. The limit of $\Lren$ for $T \to T_c^+$ is a number close to $1/2$, which, interestingly, is the value that is obtained analytically for $N \to \infty$ in $1+1$ spacetime dimensions~\cite{Gross_Witten_Wadia}. For $T>T_c$, the renormalized Polyakov loop is at first growing with the temperature (in the regime in which the plasma is most strongly coupled), it overshoots the value $1$ at temperatures around $3T_c$, reaches a maximum, and then eventually starts decreasing, in agreement with the perturbative predictions~\cite{perturbative_Polyakov_loop}.
\item In the deconfined phase, for temperatures up to approximately $3T_c$ or $4T_c$, the logarithm of the renormalized Polyakov loop (in the renormalization scheme that we considered here) is described well by the sum of a term inversely proportional to the square of the temperature, plus a constant.
\item The finite-temperature behavior of gauge theories based on different $\SU(N)$ gauge groups appears to be qualitatively and quantitatively very similar (confirming previous studies both in $3+1$~\cite{largeN3+1D} and in $2+1$ dimensions~{\cite{largeN2+1D}}). The precision and accuracy limits in this study do not allow us to extract a reliable estimate of the (small) differences between the various groups.
\end{enumerate}

In conclusion, our study shows that, in the deconfined phase, the Polyakov loop satisfies Casimir scaling, and is only mildly dependent on the number of colors $N$. The independence on the rank of the gauge group (which has also been observed for the equation of state per gluon d.o.f.~{\cite{largeN3+1D}}) supports analytical approaches based on the large-$N$ limit, including, in particular, holographic computations. Our results for the renormalized Polyakov loop show that this quantity interpolates between a regime (possibly dominated by contributions of non-perturbative nature) in which it is increasing with $T$, and one in which it tends to the perturbative prediction, and decreases with the temperature, approaching $1$ from above in the weak-coupling limit for $T \to \infty$.

For the future, we plan to extend the present study of large-$N$ gauge theories at finite temperature, by looking at other observables, which could potentially reveal a stronger dependence on the rank of the gauge group. Of particular phenomenological interest are transport and diffusion coefficients---see, e.g., ref.~\cite{Meyer:2011gj} for a review.

\vskip1.0cm {\bf Acknowledgements.}

This work is supported by the Academy of Finland, project 1134018, and in part by the National Science Foundation under Grant No. PHY11-25915. A.M. acknowledges support from the Magnus Ehrnrooth Foundation. M.P. gratefully acknowledges the Kavli Institute for Theoretical Physics in Santa Barbara, USA, for support and hospitality during the ``Novel Numerical Methods for Strongly Coupled Quantum Field Theory and Quantum Gravity'' program, during which part of this work was done. The simulations were performed at the Finnish IT Center for Science (CSC), Espoo, Finland. We thank F.~Gliozzi, O.~Kaczmarek, M.~Laine, D.~N\'ogr\'adi, R.~Pisarski, K.~Tuominen and M.~Veps\"al\"ainen for helpful comments and discussions.

\appendix{}
\vskip 0.5cm
\section{Irreducible representations of the algebra of generators of $\SU(N)$}
\label{sec:appendix}
\renewcommand{\theequation}{A.\arabic{equation}}
\setcounter{equation}{0}

In the following, we discuss the classification of irreducible representations of the algebra of generators of a generic special unitary group of degree $N$. Further details can be found, e.g., in ref.~\cite{Iachello_book}. 

A generic irreducible representation of the algebra of generators of $\SU(N)$ can be labelled by $N-1$ non-negative integers $\lambda_1, \lambda_2, \lambda_3, \dots , \lambda_{N-1}$, with:
\eq
\lambda_1 \ge \lambda_2 \ge \lambda_3 \ge \dots \ge \lambda_{N-1} \ge 0.
\en
The $[\lambda_1, \lambda_2, \lambda_3, \dots , \lambda_{N-1}]$ sequence can be uniquely associated to a Young diagram with rows of lengths $\lambda_1, \dots , \lambda_{N-1}$. An alternative way to identify an irreducible representation is in terms of its canonical label $(m_1, m_2, \dots, m_{N-1})$, where the $m_i$'s represent the differences in lengths of subsequent rows in the corresponding Young diagram: $m_i=\lambda_{i+1}-\lambda_i$ for $i<N-1$, and $m_{N-1}=\lambda_{N-1}$.

Particularly interesting irreducible representations of $\SU(N)$ include the fundamental one $[1,0,0, \dots ,0]$, of dimension $N$, the trivial one $[0,0,0, \dots ,0]$ of dimension $1$, and the adjoint one $[2,1,1, \dots, 1]$, of dimension $N^2-1$. 

More in general, the dimension of an irreducible representation is given by the formula:
\eq
\prod_{i=1}^{N-1} \prod_{j=i+1}^{N} \frac{l_i-l_j}{l_i^0-l_j^0}, \;\;\; l_i=\lambda_i + N - i, \;\;\; l_i^0= N - i,
\en
with $\lambda_N=0$ or, equivalently:
\eq
\frac{1}{\mathcal{N}_N} \prod_{l=1}^{N-1} \prod_{i=1}^{N-l} \sum_{k=i}^{i+l-1} (m_k + 1 ), \;\;\; \mbox{with:} \;\;\; \mathcal{N}_N=\prod_{t=1}^{N-1} (t!).
\en
A common way to denote irreducible representations is via their dimension; note, however, that this may be ambiguous (except for $\SU(2)$), since in general there can be inequivalent irreducible representations of the same dimension. For example, $\SU(4)$ has three inequivalent irreducible representations of dimension $20$, which can be denoted as $\mathbf{20}$, $\mathbf{20^\prime}$ and $\mathbf{20^{\prime\prime}}$. In such cases, our convention is to use the notation with the least primes for the representation with the smallest $\lambda_i$ for the minimum value of $i$ (namely, for the representation described by a Young diagram with the smallest number of boxes in the top row, or in the highest row which is different from the other representations of the same dimension). So, for instance, for $\SU(3)$ the $[3,1]$ irreducible representation is denoted as $\mathbf{15}$, while the $[4,0]$ will be denoted as $\mathbf{15^\prime}$; for $\SU(4)$, the $[2,1,0]$ is denoted as $\mathbf{20}$, the $[2,2,0]$ is denoted as $\mathbf{20^\prime}$, and the $[3,0]$ is denoted as $\mathbf{20^{\prime\prime}}$.

The $N$-ality of an $\SU(N)$ representation defines its transformation properties under the center of the group, $\Z_N$, and is given by the total number of boxes appearing in the Young diagram, modulo $N$. Representations of vanishing $N$-ality (such as the trivial representation and the adjoint one) are blind to the action of the transformations in the group center.

Given an irreducible representation $r=[\lambda_1,\lambda_2, \dots, \lambda_{N-1}]$, its conjugate representation is: $\bar{r}=[\lambda_1, \lambda_1-\lambda_{N-1}, \lambda_1-\lambda_{N-2}, \dots , \lambda_1-\lambda_2]$, so that its Young diagram is obtained by fitting the diagram of the representation $r$ in a rectangle of $N$ rows and $\lambda_1$ columns, removing all the boxes belonging to the Young diagram of $r$, and turning the diagram with the remaining boxes by an angle $\pi$. 

Obviously, two mutually conjugated representations have the same dimension, and, given that their respective characters are obtained from each other by complex conjugation, we only include one of them in our lists of irreducible representations. It is most natural to use the ``barred'' notation for the representation with the Young diagram with more boxes, so that, for example the $[1,0]$ representation of $\SU(3)$ is denoted as $\mathbf{3}$, while its conjugate representation $[1,1]$ is denoted as $\mathbf{\bar{3}}$.

Representations which are self-conjugate have real characters; in particular, this is always the case for the trivial and for the adjoint representations. Also, note that, for a self-conjugate irreducible representation, the canonical label is a palindrome.

In order to discuss the large-$N$ scaling of the size and quadratic Casimir of an irreducible representation $r$, it is convenient to introduce the non-negative integers $l$ and $m$, which represent the minimum number of fundamental and anti-fundamental factors from which the representation $r$ can be constructed (by tensor products). $l$ and $m$ can be easily obtained from the Young diagram of $r$: $l$ is given by the sum of the number of boxes in all columns of length not larger than $N/2$, while $m$ is given by the sum of the number of missing boxes in all columns of length larger than $N/2$. The $N$-ality of a representation is given by $(l-m)$~modulo~$N$. In the large-$N$ limit, it is possible to show~\cite{Gross_Taylor} that characters of different representations only depend on $l$ and $m$, and that, although the dimension of the representation $r$ grows like $N^{l+m}$, the eigenvalue of the quadratic Casimir is linear in $N$:
\begin{equation}
\langle C_2 \rangle_r = \frac{N}{2} \left[ l+m + O(1/N) \right].
\end{equation}

For $\SU(2)$, \emph{all} irreducible representations are self-conjugate. The Young diagram of a generic irreducible representation of spin $j=n/2$ consists of one horizontal row of $n$ boxes; bosonic representations correspond to even values of $n$, and have vanishing $N$-ality, while fermionic representations correspond to odd values of $n$, and their $N$-ality is $1$. The associated canonical label is $(n)$ (with $l=n$, $m=0$), the dimension is $n+1$, the eigenvalue of the quadratic Casimir (defined according to our conventions) is $\langle C_2 \rangle = n(n+2)/4$, and its ratio with respect to the fundamental representation is $d=n(n+2)/3$.

For larger $\SU(N)$ groups (up to $N=8$), the lowest irreducible representations are listed in tables~\ref{SU3_irreps}--\ref{SU8_irreps}.

\begin{table}[h]
\centering
\phantom{-------}
\begin{tabular}{|c|c|c|c|c|c|c|}  
\hline
Young diagram & $N$-ality & canonical label & dimension & notes & $\langle C_2 \rangle$ & $d$ \\
\hline
 $\tiny{\yng(1)}$    & $1$ & $(1,0)$ &  $\mathbf{3}$ & fundamental & $4/3$ & $1$ \\
 $\tiny{\yng(2)}$    & $2$ & $(2,0)$ &  $\mathbf{6}$ & & $10/3$  & $5/2$ \\
 $\tiny{\yng(2,1)}$  & $0$ & $(1,1)$ &  $\mathbf{8}$ & adjoint & $3$ & $9/4$ \\
 $\tiny{\yng(3)}$    & $0$ & $(3,0)$ & $\mathbf{10}$ & &  $6$ & $9/2$ \\
 $\tiny{\yng(3,1)}$  & $1$ & $(2,1)$ & $\mathbf{15}$ & & $16/3$  & $4$ \\
 $\tiny{\yng(4)}$    & $1$ & $(4,0)$ & $\mathbf{15^\prime}$ & & $28/3$  & $7$ \\
 $\tiny{\yng(5)}$    & $2$ & $(5,0)$ & $\mathbf{21}$ & &  $40/3$ & $10$ \\
 $\tiny{\yng(4,1)}$  & $2$ & $(3,1)$ & $\mathbf{24}$ & \phantom{$\tiny{\yng(1,1,1)}$} &  $25/3$ & $25/4$ \\
 $\tiny{\yng(4,2)}$  & $0$ & $(2,2)$ & $\mathbf{27}$ & self-conjugate & $8$ & $6$ \\
 $\tiny{\yng(6)}$    & $0$ & $(6,0)$ & $\mathbf{28}$ & & $18$  & $27/2$ \\
 $\tiny{\yng(5,1)}$  & $0$ & $(4,1)$ & $\mathbf{35}$ & & $12$  & $9$ \\
 $\tiny{\yng(7)}$    & $1$ & $(7,0)$ & $\mathbf{36}$ & & $70/3$  & $35/2$ \\
\hline                 
\end{tabular}
\phantom{-------}
\caption{The irreducible representations of the $\SU(3)$ gauge group studied in this work. For this group, the integers $l$ and $m$ of each representation are respectively equal to the first and second index in the canonical label.}
\label{SU3_irreps}
\end{table}

\begin{table}[h]
\centering
\phantom{-------}
\begin{tabular}{|c|c|c|c|c|c|c|c|c|}  
\hline
Young diagram & $N$-ality & canonical label & dimension & $l$ & $m$ & notes & $\langle C_2 \rangle$ & $d$ \\
\hline
 $\tiny{\yng(1)}$      & $1$ & $(1,0,0)$ &  $\mathbf{4}$ & $1$ & $0$ & fundamental & $15/8$ & $1$ \\
 $\tiny{\yng(1,1)}$    & $2$ & $(0,1,0)$ &  $\mathbf{6}$ & $2$ & $0$  & self-conjugate & $5/2$ & $4/3$ \\
 $\tiny{\yng(2)}$      & $2$ & $(2,0,0)$ & $\mathbf{10}$ & $2$ & $0$  & \phantom{$\tiny{\yng(1,1)}$}  & $9/2$ & $12/5$ \\
 $\tiny{\yng(2,1,1)}$   & $0$ & $(1,0,1)$ & $\mathbf{15}$ & $1$ & $1$  & adjoint & $4$ & $32/15$ \\
 $\tiny{\yng(2,1)}$    & $3$ & $(1,1,0)$ & $\mathbf{20}$ & $3$ & $0$  & \phantom{$\tiny{\yng(1,1,1)}$} & $39/8$ & $13/5$ \\
 $\tiny{\yng(2,2)}$    & $0$ & $(0,2,0)$ & $\mathbf{20^\prime}$ & $4$ & $0$  & self-conjugate & $6$ & $16/5$ \\
 $\tiny{\yng(3)}$      & $3$ & $(3,0,0)$ & $\mathbf{20^{\prime\prime}}$ & $3$ & $0$  &  & $63/8$ & $21/5$ \\
 $\tiny{\yng(4)}$      & $0$ & $(4,0,0)$ & $\mathbf{35}$ & $4$ & $0$  & \phantom{$\tiny{\yng(1,1)}$} & $12$ & $32/5$ \\
 $\tiny{\yng(3,1,1)}$  & $1$ &  $(2,0,1)$ & $\mathbf{36}$ & $2$ & $1$  &  & $55/8$ & $11/3$ \\
 $\tiny{\yng(3,1)}$    & $0$ &  $(2,1,0)$ & $\mathbf{45}$ & $4$ & $0$  & \phantom{$\tiny{\yng(1,1,1)}$} & $8$ & $64/15$ \\
 $\tiny{\yng(3,3)}$    & $2$ & $(0,3,0)$ & $\mathbf{50}$ & $6$ & $0$  & self-conjugate & $21/2$ & $28/5$ \\
 $\tiny{\yng(5)}$      & $1$ & $(5,0,0)$ & $\mathbf{56}$ & $5$ & $0$  &  & $135/8$ & $9$ \\
\hline
\end{tabular}
\phantom{-------}
\caption{Same as in table~\ref{SU3_irreps} (with the addition of the $l$ and $m$ indices), but for the $\SU(4)$ gauge group.}
\label{SU4_irreps}
\end{table}

\begin{table}[h]
\centering
\phantom{-------}
\begin{tabular}{|c|c|c|c|c|c|c|c|c|}  
\hline
Young diagram & $N$-ality & canonical label & dimension & $l$ & $m$ & notes & $\langle C_2 \rangle$ & $d$ \\
\hline
 $\tiny{\yng(1)}$       & $1$ & $(1,0,0,0)$ &   $\mathbf{5}$ & $1$ & $0$ & fundamental & $12/5$ & $1$ \\
 $\tiny{\yng(1,1)}$     & $2$ & $(0,1,0,0)$ &  $\mathbf{10}$ & $2$ & $0$ &  & $18/5$ & $3/2$ \\
 $\tiny{\yng(2)}$       & $2$ & $(2,0,0,0)$ &  $\mathbf{15}$ & $2$ & $0$ &  & $28/5$ & $7/3$ \\
 $\tiny{\yng(2,1,1,1)}$ & $0$ & $(1,0,0,1)$ &  $\mathbf{24}$ & $1$ & $1$ & adjoint  & $5$ & $25/12$ \\
 $\tiny{\yng(3)}$       & $3$ & $(3,0,0,0)$ &  $\mathbf{35}$ & $3$ & $0$ &   & $48/5$ & $4$ \\
 $\tiny{\yng(2,1)}$     & $3$ & $(1,1,0,0)$ &  $\mathbf{40}$ & $3$ & $0$ & \phantom{$\tiny{\yng(1,1,1)}$}  & $33/5$ & $11/4$ \\
 $\tiny{\yng(2,1,1)}$   & $4$ & $(1,0,1,0)$ &  $\mathbf{45}$ & $1$ & $2$ & \phantom{$\tiny{\yng(1,1,1,1)}$} & $32/5$ & $8/3$ \\
 $\tiny{\yng(2,2)}$     & $4$ & $(0,2,0,0)$ &  $\mathbf{50}$ & $4$ & $0$ & \phantom{$\tiny{\yng(1,1,1)}$} & $42/5$ & $7/2$ \\
 $\tiny{\yng(3,1,1,1)}$ & $1$ & $(2,0,0,1)$ &  $\mathbf{70}$ & $2$ & $1$ & \phantom{$\tiny{\yng(1,1,1,1,1)}$} & $42/5$ & $7/2$ \\
 $\tiny{\yng(4)}$       & $4$ & $(4,0,0,0)$ &  $\mathbf{70^\prime}$ & $4$ & $0$ &  & $72/5$ & $6$  \\
 \phantom{$\tiny{\yng(1,1,1,1)}$} $\tiny{\yng(2,2,1)}$ \phantom{$\tiny{\yng(1,1,1,1)}$} & $0$ & $(0,1,1,0)$ &  $\mathbf{75}$ & $2$ & $2$ & self-conjugate  & $8$ & $10/3$ \\
 $\tiny{\yng(3,1)}$     & $4$ & $(2,1,0,0)$ & $\mathbf{105}$ & $4$ & $0$ & \phantom{$\tiny{\yng(1,1,1)}$}  & $52/5$ & $13/3$ \\
\hline
\end{tabular}
\phantom{-------}
\caption{Same as in table~\ref{SU4_irreps}, but for the $\SU(5)$ gauge group.}
\label{SU5_irreps}
\end{table}

\begin{table}[h]
\centering
\phantom{-------}
\begin{tabular}{|c|c|c|c|c|c|c|c|c|}  
\hline
Young diagram & $N$-ality & canonical label & dimension & $l$ & $m$ & notes & $\langle C_2 \rangle$ & $d$ \\
\hline
 $\tiny{\yng(1)}$         & $1$ & $(1,0,0,0,0)$ &   $\mathbf{6}$ & $1$ & $0$ & fundamental & $35/12$ & $1$ \\
 $\tiny{\yng(1,1)}$       & $2$ & $(0,1,0,0,0)$ &  $\mathbf{15}$ & $2$ & $0$ & \phantom{$\tiny{\yng(1,1,1)}$}  & $14/3$ & $8/5$ \\
 $\tiny{\yng(1,1,1)}$     & $3$ & $(0,0,1,0,0)$ &  $\mathbf{20}$ & $3$ & $0$ & self-conjugate  & $21/4$ & $9/5$ \\
 $\tiny{\yng(2)}$         & $2$ & $(2,0,0,0,0)$ &  $\mathbf{21}$ & $2$ & $0$ &  & $20/3$ & $16/7$ \\
 $\tiny{\yng(2,1,1,1,1)}$ & $0$ & $(1,0,0,0,1)$ &  $\mathbf{35}$ & $1$ & $1$ & adjoint  & $6$ & $72/35$ \\
 $\tiny{\yng(3)}$         & $3$ & $(3,0,0,0,0)$ &  $\mathbf{56}$ & $3$ & $0$ &  & $45/4$ & $27/7$ \\
 $\tiny{\yng(2,1)}$       & $3$ & $(1,1,0,0,0)$ &  $\mathbf{70}$ & $3$ & $0$ &  & $33/4$ & $99/35$ \\
 $\tiny{\yng(2,1,1,1)}$   & $5$ & $(1,0,0,1,0)$ &  $\mathbf{84}$ & $1$ & $2$ & \phantom{$\tiny{\yng(1,1,1,1,1)}$}  & $95/12$ & $19/7$ \\
 $\tiny{\yng(2,1,1)}$     & $4$ & $(1,0,1,0,0)$ & $\mathbf{105}$ & $4$ & $0$ &  & $26/3$ & $104/35$ \\
 $\tiny{\yng(2,2)}$       & $4$ & $(0,2,0,0,0)$ &  $\mathbf{105^\prime}$ & $4$ & $0$ & \phantom{$\tiny{\yng(1,1,1)}$}  & $32/3$ & $128/35$ \\
 $\tiny{\yng(3,1,1,1,1)}$ & $1$ & $(2,0,0,0,1)$ & $\mathbf{120}$ & $2$ & $1$ &  & $119/12$ & $17/5$ \\
 $\tiny{\yng(4)}$         & $4$ & $(4,0,0,0,0)$ & $\mathbf{126}$ & $4$ & $0$ &  & $50/3$ & $40/7$ \\
\hline
\end{tabular}
\phantom{-------}
\caption{Same as in table~\ref{SU4_irreps}, but for the $\SU(6)$ gauge group.}
\label{SU6_irreps}
\end{table}

\begin{table}[h]
\centering
\phantom{-------}
\begin{tabular}{|c|c|c|c|c|c|c|c|c|}  
\hline
Young diagram & $N$-ality & canonical label & dimension & $l$ & $m$ & notes & $\langle C_2 \rangle$ & $d$ \\
\hline
 $\tiny{\yng(1)}$             	& $1$ & $(1,0,0,0,0,0)$ & $\mathbf{7}$ 	& $1$ & $0$ & fundamental & $24/7$ & $1$ \\
 $\tiny{\yng(1,1)}$           	& $2$ & $(0,1,0,0,0,0)$ & $\mathbf{21}$ 	& $2$ & $0$ &  \phantom{$\tiny{\yng(1,1)}$} & $40/7$ & $5/3$ \\
 $\tiny{\yng(2)}$             	& $2$ & $(2,0,0,0,0,0)$ & $\mathbf{28}$ 	& $2$ & $0$ &  & $54/7$ & $9/4$ \\
 $\tiny{\yng(1,1,1)}$         	& $3$ & $(0,0,1,0,0,0)$ & $\mathbf{35}$ 	& $3$ & $0$ &  \phantom{$\tiny{\yng(1,1,1,1)}$} & $48/7$ & $2$ \\
  $\tiny{\yng(2,1,1,1,1,1)}$    & $0$ & $(1,0,0,0,0,1)$ & $\mathbf{48}$ 	& $1$ & $1$ &  adjoint & \phantom{$\tiny{\yng(1,1,1,1,1,1)}$} $\!\!\!\!\! 7$ & $49/24$ \\
 $\tiny{\yng(3)}$       	& $3$ & $(3,0,0,0,0,0)$ & $\mathbf{84}$ 	& $3$ & $0$ &  & $90/7$ & $15/4$ \\
 $\tiny{\yng(2,1)}$             & $3$ & $(1,1,0,0,0,0)$ & $\mathbf{112}$ 	& $3$ & $0$ &  \phantom{$\tiny{\yng(1,1,1)}$} & $69/7$ & $23/8$ \\
 $\tiny{\yng(2,1,1,1,1)}$       & $6$ & $(1,0,0,0,1,0)$ & $\mathbf{140}$ 	& $1$ & $2$ &  \phantom{$\tiny{\yng(1,1,1,1,1)}$} & $66/7$ & $11/4$ \\
 $\tiny{\yng(3,1,1,1,1,1)}$   	& $1$ & $(2,0,0,0,0,1)$ & $\mathbf{189}$ 	& $2$ & $1$ &  \phantom{$\tiny{\yng(1,1,1,1,1,1,1)}$} & $80/7$ & $10/3$ \\
 $\tiny{\yng(2,2)}$ 		& $4$ & $(0,2,0,0,0,0)$ & $\mathbf{196}$ 	& $4$ & $0$ &  \phantom{$\tiny{\yng(1,1)}$} & $90/7$ & $15/4$ \\
 $\tiny{\yng(2,1,1)}$           & $4$ & $(1,0,1,0,0,0)$ & $\mathbf{210}$ 	& $4$ & $0$ &  \phantom{$\tiny{\yng(1,1,1,1)}$} & $76/7$ & $19/6$ \\
 $\tiny{\yng(4)}$           	& $4$ & $(4,0,0,0,0,0)$ & $\mathbf{210'}$ 	& $4$ & $0$ &  & $132/7$ & $11/2$ \\
\hline
\end{tabular}
\phantom{-------}
\caption{Same as in table~\ref{SU4_irreps}, but for the $\SU(7)$ gauge group.}
\label{SU7_irreps}
\end{table}

\begin{table}[h]
\centering
\phantom{-------}
\begin{tabular}{|c|c|c|c|c|c|c|c|c|}  
\hline
Young diagram & $N$-ality & canonical label & dimension & $l$ & $m$ & notes & $\langle C_2 \rangle$ & $d$ \\
\hline
 $\tiny{\yng(1)}$             & $1$ & $(1,0,0,0,0,0,0)$ &   $\mathbf{8}$ & $1$ & $0$ & fundamental & $63/16$ & $1$ \\
 $\tiny{\yng(1,1)}$           & $2$ & $(0,1,0,0,0,0,0)$ &  $\mathbf{28}$ & $2$ & $0$  & \phantom{$\tiny{\yng(1,1,1)}$}  & $27/4$ & $12/7$ \\
 $\tiny{\yng(2)}$             & $2$ & $(2,0,0,0,0,0,0)$ &  $\mathbf{36}$ & $2$ & $0$  &  & $35/4$ & $20/9$ \\
 $\tiny{\yng(1,1,1)}$         & $3$ & $(0,0,1,0,0,0,0)$ &  $\mathbf{56}$ & $3$ & $0$  & \phantom{$\tiny{\yng(1,1,1,1)}$} & $135/16$ & $15/7$ \\
 \phantom{$\tiny{\yng(1,1,1,1,1,1,1,1)}$} $\tiny{\yng(2,1,1,1,1,1,1)}$ \phantom{$\tiny{\yng(1,1,1,1,1,1,1,1)}$} & $0$ & $(1,0,0,0,0,0,1)$ &  $\mathbf{63}$ & $1$ & $1$  & adjoint  & $8$ & $128/63$ \\
 $\tiny{\yng(1,1,1,1)}$       & $4$ & $(0,0,0,1,0,0,0)$ &  $\mathbf{70}$ & $4$ & $0$  & self-conjugate  & $9$ & $16/7$ \\
 $\tiny{\yng(3)}$             & $3$ & $(3,0,0,0,0,0,0)$ & $\mathbf{120}$ & $3$ & $0$  &  & $231/16$ & $11/3$ \\
 $\tiny{\yng(2,1)}$           & $3$ & $(1,1,0,0,0,0,0)$ & $\mathbf{168}$ & $3$ & $0$  & \phantom{$\tiny{\yng(1,1,1)}$}  & $183/16$ & $61/21$ \\
 $\tiny{\yng(2,1,1,1,1,1)}$   & $7$ & $(1,0,0,0,0,1,0)$ & $\mathbf{216}$ & $1$ & $2$  & \phantom{$\tiny{\yng(1,1,1,1,1,1,1)}$}  & $175/16$ & $25/9$ \\
 $\tiny{\yng(3,1,1,1,1,1,1)}$ & $1$ & $(2,0,0,0,0,0,1)$ & $\mathbf{280}$ & $2$ & $1$  & \phantom{$\tiny{\yng(1,1,1,1,1,1,1,1)}$} & $207/16$ & $23/7$ \\
 $\tiny{\yng(4)}$             & $4$ & $(4,0,0,0,0,0,0)$ & $\mathbf{330}$ & $4$ & $0$  &  & $21$ & $16/3$ \\
 $\tiny{\yng(2,2)}$           & $4$ & $(0,2,0,0,0,0,0)$ & $\mathbf{336}$ & $4$ & $0$  & \phantom{$\tiny{\yng(1,1,1)}$}  & $15$ & $80/21$ \\
\hline
\end{tabular}
\phantom{-------}
\caption{Same as in table~\ref{SU4_irreps}, but for the $\SU(8)$ gauge group.}
\label{SU8_irreps}
\end{table}

Generically, the eigenvalues of a group element $g$ in the fundamental representation of $\SU(N)$ lie on the unit circle in the complex plane, and their product is $1$:
\eq
g_f = U \cdot \mbox{diag}( e^{i \alpha_1} , e^{i \alpha_2} , e^{i \alpha_3} , \dots , e^{i \alpha_N}) \cdot U^\dagger, \;\;\; \mbox{with: } \sum_{i=1}^N \alpha_i = 0 \mbox{ mod } 2 \pi. 
\en
Knowing the eigenvalues of $g_{\mbox\tiny{f}}$, it is possible to calculate explicitly the character of $g$ in \emph{any} irreducible representation $r=[\lambda_1,\lambda_2,\dots,\lambda_{N-1}]$ by means of the Weyl formula~\cite{Weyl_formula}:
\eq
\Tr g_r = \frac{ \det F(\vec{\lambda})}{ \det F(\vec{0})},
\en 
where $F(\vec{\lambda})$ is an $N \times N$ matrix with entries defined as: $F_{kl}(\vec{\lambda}) = \exp\left[ i \left( N+\lambda_l-l \right) \alpha_k \right]$, with $\lambda_N=0$, and $e^{i \alpha_1}$, $e^{i \alpha_2}$, $\dots$ $e^{i \alpha_N}$ are the eigenvalues of $g$ in the fundamental representation.

In many cases, however, the characters in high-dimensional irreducible representations can be more expediently calculated, using the laws of representation composition encoded in Young calculus, and using the well-known fact that the character in a representation which can be expressed as the tensor sum (product) of two representations is equal to the sum (product) of the characters in the summand (factor) representations.

\subsection{Casimir operators}
\label{subsec:Casimir_operator_definition}

A Casimir operator of a Lie algebra $g$ is a homogeneous polynomial of order $p$, lying in the enveloping algebra of $g$, $T(g)$, and commuting with all elements of $g$. Given a Casimir operator $C_p$, any product of it by an arbitrary scalar factor $aC_p$, as well as any integer power of it $C_p^q$, are also Casimir operators; however, the number of \emph{independent} Casimir operators of a given algebra $g$ is equal to the rank $l$ of the algebra. In particular, the algebra of generators of the special unitary group $\SU(N)$ has $N-1$ independent Casimir operators $C_2$, $C_3$, \dots $C_N$, whose eigenvalues $\langle C_p \rangle$~can be used to classify the irreducible representations of the algebra. 

Explicit expressions for the $C_p$'s can be obtained as follows. Starting from a basis $\left\{ E_{i,j} \right\}_{i,j=1\dots N}$ of generators of $\U(N)$:
\begin{equation}
\label{u_N_algebra}
\left[ E_{a,b}, E_{c,d}\right] = \delta_{b,c} E_{a,d} - \delta_{a,d} E_{c,b},
\end{equation}
introduce a basis for the algebra of generators, denoted as $\left\{ \tilde{E}_{i,j} \right\}$ (where both $i$ and $j$ run from $1$ to $N$, but the element $\tilde{E}_{N,N}$ element is not defined), through:
\begin{equation}
\label{tildeE_definitions}
\tilde{E}_{i,j} = \left\{ 
\begin{array}{ll}
E_{i,j} & \mbox{if } i\ne j, \\
E_{i,i} -\frac{1}{N}\sum_{k=1}^{N} E_{k,k} & \mbox{if } i=j. \\
\end{array}
\right.
\end{equation}
For the generators of $\SU(N)$, the Casimir operator of order $p$ can then be defined as:
\begin{equation}
\label{C_p_definition}
C_p = \frac{1}{p} \sum_{i_1, i_2, \dots i_p=1}^{N} \tilde{E}_{i_1 i_2} \tilde{E}_{i_2 i_3} \dots \tilde{E}_{i_{p-1} i_p} \tilde{E}_{i_p i_1} .
\end{equation}
Note that, by construction, the linear Casimir operator $C_1$ is identically vanishing on the algebra of generators of $\SU(N)$, as they are all traceless.

The eigenvalue of $C_p$ in the generic irreducible representation labelled by $[\lambda_1, \lambda_2, \dots \lambda_{N-1}]$ can be obtained in the following way (taking $\lambda_N=0$)~\cite{Perelomov_Popov_SovJNuclPhys_3_676_1966}:

\begin{enumerate}
\item define $\lambda=\sum_{i=1}^{N} \lambda_i$;
\item define $m_i=\lambda_i-\lambda/N$ for all $i=1$, $2$, \dots $N$;
\item define $\rho_i=N-i$ and $l_i=m_i+\rho_i$ for all $i=1$, $2$, \dots $N$;
\item for all $k \ge 2$, construct the quantities: $S_k=\sum_{i=1}^N \left( l_i^k - \rho_i^k \right)$;
\item for all $k \ge 2$, define the coefficients: $a_k=\sum_{j=1}^{k-1} \frac{(k-1)!}{j!(k-j)!} S_j$;
\item construct the function: $\varphi(z)=\sum_{k=2}^{\infty} a_k z^k$;
\item calculate the $B_p$ coefficients from the following Taylor expansion around $z=0$:
\begin{equation}
\label{Bp_from_Taylor_expansion}
\frac{ 1 - \exp\left[ -\varphi(z) \right] }{z} = \sum_{p=0}^{\infty} B_p z^p
\end{equation}
(note that $B_0=0$);
\item compute the eigenvalue of $C_p$ from the formula:
\begin{equation}
\label{Cp_eigenvalue}
\langle C_p \rangle = \frac{B_p - N B_{p-1}}{p}.
\end{equation}
\end{enumerate}

This gives, in particular, the following relations:
\begin{equation}
\label{C2}
\langle C_2 \rangle = \frac{S_2}{2},
\end{equation}
\begin{equation}
\label{C3}
\langle C_3 \rangle = \frac{1}{3} \left[ S_3 
  + \left(\frac{3}{2}-N \right)S_2 \right]\!,
\end{equation}
\begin{equation}
\label{C4}
\langle C_4 \rangle = \frac{1}{4} \left[ S_4
  + \left( 2-N \right)S_3
  + \left( 2- \frac{3}{2}N \right)S_2 \right]\!,
\end{equation}
\begin{equation}
\label{C5}
\langle C_5 \rangle = \frac{1}{5} \left[ S_5
  + \left( \frac{5}{2} -N \right)S_4
  + \left( \frac{10}{3} -2N \right)S_3
  + \left( \frac{5}{2} -2N \right)S_2
  - \frac{1}{2}S_2^2 \right]\!,
\end{equation}
\begin{eqnarray}
\label{C6}
\langle C_6 \rangle &=& \frac{1}{6} \left[ S_6
  + \left( 3-N \right)S_5
  + \left( 5- \frac{5}{2}N\right)S_4
  + \left( 5- \frac{10}{3}N\right)S_3
  + \left( 3- \frac{5}{2}N\right)S_2 \right. \nonumber \\
& & \left. - S_2S_3
  + \left( \frac{N}{2} - \frac{3}{2}\right)S_2^2 \right]\!,
\end{eqnarray}
\begin{eqnarray}
\label{C7}
\langle C_7 \rangle &=& \frac{1}{7} \left[ S_7
  + \left( \frac{7}{2}-N \right)S_6
  + \left( 7-3N \right)S_5
  + \left( \frac{35}{4}-5N \right)S_4
  + \left( 7-5N \right)S_3 \right.  \nonumber \\
& & \left. + \left( \frac{7}{2}-3N \right)S_2
  - S_4S_2
  - \frac{1}{2}S_3^2
  + \left( -\frac{7}{2} +N \right)S_3S_2
  + \left( -\frac{25}{8} +\frac{3}{2}N\right)S_2^2 \right] \!,
\end{eqnarray}
\begin{eqnarray}
\label{C8}
\langle C_8 \rangle &=&  \frac{1}{8} \left[S_8
  + \left( 4-N \right)S_7
  + \left( \frac{28}{3} - \frac{7}{2}N\right)S_6
  + \left( 14-7N \right)S_5
  + \left( 14-\frac{35}{4}N \right)S_4 \right. \nonumber \\
& & + \left( \frac{28}{3} -7N \right)S_3
  + \left( 4-\frac{7}{2}N \right)S_2
  - S_5S_2
  - S_4S_3
  + \left( -4+N \right)S_4S_2 \nonumber \\
& & \left. + \left( -2+\frac{N}{2} \right)S_3^2
  + \left( -\frac{25}{3} +\frac{7}{2}N \right)S_3S_2
  + \left( -\frac{11}{2} + \frac{25}{8}N \right)S_2^2
  + \frac{1}{6}S_2^3 \right]\!.
\end{eqnarray}

In turn, the equations above lead to the following expressions for the quadratic Casimir eigenvalues $\langle C_2 \rangle$:

\begin{equation}
\label{C2_for_SU2}
\langle C_2 \rangle = \frac{1}{4} \lambda_1 \left( \lambda_1 + 2 \right) \;\;\;\mbox{for $\SU(2)$},
\end{equation}
\begin{equation}
\label{C2_for_SU3}
\langle C_2 \rangle = \frac{1}{3} \left( \lambda_1^2 + 3 \lambda_1 - \lambda_1 \lambda_2 + \lambda_2^2 \right) \;\;\;\mbox{for $\SU(3)$},
\end{equation}
\begin{equation}
\label{C2_for_SU4}
\langle C_2 \rangle = \frac{1}{8} \left[ 3\lambda_1^2 
   +\lambda_2 \left( 4+3\lambda_2 \right) 
   -2\lambda_3 \left( \lambda_2 + 2 \right) 
   +3\lambda_3^2  -2 \lambda_1 \left( \lambda_2+\lambda_3-6\right)\right] \;\;\;\mbox{for $\SU(4)$}, 
\end{equation}
\begin{eqnarray}
\label{C2_for_SU5}
\langle C_2 \rangle &=& \frac{1}{5} \left[ 
  2\left( \lambda_1^2 + \lambda_2^2 + \lambda_3^2 + \lambda_4^2 \right) - \left( 5+\lambda_3 \right) \lambda_4 + \lambda_2 \left( 5-\lambda_3-\lambda_4\right) \right. \nonumber \\
 && \left. + \lambda_1\left( 10-\lambda_2-\lambda_3-\lambda_4 \right)
  \right] \;\;\;\mbox{for $\SU(5)$},
\end{eqnarray}
\begin{eqnarray}
\label{C2_for_SU6}
\langle C_2 \rangle &=& \frac{1}{12} \left[ 5 \left( \lambda_1^2 + \lambda_2^2 +  \lambda_3^2 + \lambda_4^2 + \lambda_5^2 \right) + 6 \left( \lambda_3 - \lambda_4 \right) -2 \lambda_3 \lambda_4 -2 \left( 9 + \lambda_3 + \lambda_4 \right) \lambda_5 \right. \nonumber \\
 && \left.  + 2 \lambda_2 \left( 9 - \lambda_3 - \lambda_4 - \lambda_5 \right) + 2\lambda_1 \left( 15 - \lambda_2 - \lambda_3 - \lambda_4 - \lambda_5 \right)
\right] \;\;\;\mbox{for $\SU(6)$},
\end{eqnarray}
\begin{eqnarray}
\label{C2_for_SU7}
\langle C_2 \rangle &=& \frac{1}{7} \left[ 3 \left( \lambda_1^2 + \lambda_2^2 + \lambda_3^2 + \lambda_4^2 + \lambda_5^2 + \lambda_6^2 \right) + 7 \left( 3 \lambda_1 + 2\lambda_2 + \lambda_3 - \lambda_5 -2\lambda_6 \right) - \lambda_4\lambda_5 \right. \nonumber \\
&& \left. - \lambda_3 \left( \lambda_4 + \lambda_5\right) - \left( \lambda_3 + \lambda_4 + \lambda_5 \right) \lambda_6 - \left( \lambda_3 + \lambda_4 + \lambda_5 + \lambda_6 \right) \lambda_2 \right. \nonumber \\
&& \left.  - \lambda_1 \left( \lambda_2 + \lambda_3 + \lambda_4 + \lambda_5 + \lambda_6\right)
 \right] \;\;\;\mbox{for $\SU(7)$},
\end{eqnarray}
\begin{eqnarray}
\label{C2_for_SU8}
\langle C_2 \rangle &=& \frac{1}{16} \left[ 7 \left( \lambda_1^2 + \lambda_2^2 + \lambda_3^2 + \lambda_4^2 + \lambda_5^2 + \lambda_6^2 + \lambda_7^2 \right) + 24\lambda_3 + 8 \lambda_4 -2 \lambda_3 \lambda_4 -8 \lambda_5 - 2 \lambda_3 \lambda_5 \right. \nonumber \\
&& \left.   - 2 \lambda_4 \lambda_5 -24 \lambda_6 -2 \lambda_3\lambda_6 -2 \lambda_4\lambda_6 -2 \lambda_5\lambda_6 -2 \left( 20 + \lambda_3 + \lambda_4 + \lambda_5 + \lambda_6 \right)\lambda_7  \right. \nonumber \\
&& \left.  -2\lambda_2 \left( -20 + \lambda_3 + \lambda_4 + \lambda_5 + \lambda_6 + \lambda_7 \right)  \right. \nonumber \\
&& \left.   -2\lambda_1 \left( -28 + \lambda_2 + \lambda_3 + \lambda_4 + \lambda_5 + \lambda_6 + \lambda_7 \right) \right] \;\;\;\mbox{for $\SU(8)$} .
\end{eqnarray}

Note that the Casimir operators are defined up to a multiplicative constant; with the conventions fixed by the construction above, the eigenvalue of the $\SU(N)$ quadratic Casimir operator in the fundamental representation is $(N^2-1)/(2N)$, while in the adjoint representation it is $N$.

\end{document}